\documentclass[aps,prd,twocolumn,showpcs,groupedaddress]{revtex4}

\usepackage{bm}
\newcommand{\dmes}       {\ensuremath{ D}}
\newcommand{\ad}       {\ensuremath{\bar{ D}}}
\newcommand{\dzero}     {\ensuremath{ D^0} \,}
\newcommand{\adzero}     {\ensuremath{\bar{ D}^0 \,}}
\newcommand{\dminus}     {\ensuremath{ D^-}}
\newcommand{\dst}       {\ensuremath{ D^{*} \,}}
\newcommand{\adst}       {\ensuremath{\bar{ D}^{*} \,}}

\newcommand{\adstzero}   {\ensuremath{\bar{ D}^{*0}}}
\newcommand{\dstplus}   {\ensuremath{ D^{*+}}}
\newcommand{\dstminus}  {\ensuremath{ D^{*-}}}

\newcommand{\adststzero} {\ensuremath{\bar{ D}^{**0}}}

\newcommand{\adstst}     {\ensuremath{\bar{ D}^{**}}}
\newcommand{\dststminus} {\ensuremath{ D^{**-}}}

\newcommand{\dmd}       {\ensuremath{{\Delta {m}_{d}}}}

\newcommand{\bzerod}    {\ensuremath{ B^0}}

\newcommand{\GVc}       {\ensuremath{\mathrm{GeV}/c}}
\newcommand{\GVcs}      {\ensuremath{\mathrm{GeV}/c^2}}

\newcommand{\dilu}      {\ensuremath{{\cal D}_u}} 
\newcommand{\dild}      {\ensuremath{{\cal D}_d}} 
\newcommand{\merit}     {\ensuremath{\varepsilon {\cal{D}}^2}} 
\newcommand{\doubcol} {\multicolumn{2}{c|}}

\usepackage{graphicx}
\usepackage{rotating}
\usepackage{longtable}
\usepackage{amsmath}
\usepackage{helvet}

\begin{document}

\hspace{5.2in} \mbox{FERMILAB-PUB-06/341-E}
\title{Measurement of $B_d$ mixing using opposite-side flavor tagging}


%
\author{                                                                      
V.M.~Abazov,$^{35}$                                                           
B.~Abbott,$^{75}$                                                             
M.~Abolins,$^{65}$                                                            
B.S.~Acharya,$^{28}$                                                          
M.~Adams,$^{51}$                                                              
T.~Adams,$^{49}$                                                              
M.~Agelou,$^{17}$                                                             
E.~Aguilo,$^{5}$                                                              
S.H.~Ahn,$^{30}$                                                              
M.~Ahsan,$^{59}$
% ...
\\                                                                            
\vskip 0.30cm                                                                 
\centerline{(D\O\ Collaboration)}                                             
\vskip 0.30cm                                                                 
}                                                                             
\affiliation{                                                                 
\centerline{$^{1}$Universidad de Buenos Aires, Buenos Aires, Argentina}       
\centerline{$^{2}$LAFEX, Centro Brasileiro de Pesquisas F{\'\i}sicas,         
                  Rio de Janeiro, Brazil}                                     
\centerline{$^{3}$Universidade do Estado do Rio de Janeiro,                   
                  Rio de Janeiro, Brazil}                                     
\centerline{$^{4}$Instituto de F\'{\i}sica Te\'orica, Universidade            
                  Estadual Paulista, S\~ao Paulo, Brazil}                     
\centerline{$^{5}$University of Alberta, Edmonton, Alberta, Canada,           
                  Simon Fraser University, Burnaby, British Columbia, Canada,}
\centerline{York University, Toronto, Ontario, Canada, and                    
                  McGill University, Montreal, Quebec, Canada}                
\centerline{$^{6}$University of Science and Technology of China, Hefei,       
                  People's Republic of China}                                 
\centerline{$^{7}$Universidad de los Andes, Bogot\'{a}, Colombia}             
\centerline{$^{8}$Center for Particle Physics, Charles University,            
                  Prague, Czech Republic}                                     
\centerline{$^{9}$Czech Technical University, Prague, Czech Republic}         
\centerline{$^{10}$Center for Particle Physics, Institute of Physics,         
                   Academy of Sciences of the Czech Republic,                 
                   Prague, Czech Republic}                                    
\centerline{$^{11}$Universidad San Francisco de Quito, Quito, Ecuador}        
\centerline{$^{12}$Laboratoire de Physique Corpusculaire, IN2P3-CNRS,         
                   Universit\'e Blaise Pascal, Clermont-Ferrand, France}      
\centerline{$^{13}$Laboratoire de Physique Subatomique et de Cosmologie,      
                   IN2P3-CNRS, Universite de Grenoble 1, Grenoble, France}    
\centerline{$^{14}$CPPM, IN2P3-CNRS, Universit\'e de la M\'editerran\'ee,     
                   Marseille, France}                                         
\centerline{$^{15}$IN2P3-CNRS, Laboratoire de l'Acc\'el\'erateur              
                   Lin\'eaire, Orsay, France}                                 
\centerline{$^{16}$LPNHE, IN2P3-CNRS, Universit\'es Paris VI and VII,         
                   Paris, France}                                             
\centerline{$^{17}$DAPNIA/Service de Physique des Particules, CEA, Saclay,    
                   France}                                                    
\centerline{$^{18}$IPHC, IN2P3-CNRS, Universit\'e Louis Pasteur, Strasbourg,  
                    France, and Universit\'e de Haute Alsace,                 
                    Mulhouse, France}                                         
\centerline{$^{19}$Institut de Physique Nucl\'eaire de Lyon, IN2P3-CNRS,      
                   Universit\'e Claude Bernard, Villeurbanne, France}         
\centerline{$^{20}$III. Physikalisches Institut A, RWTH Aachen,               
                   Aachen, Germany}                                           
\centerline{$^{21}$Physikalisches Institut, Universit{\"a}t Bonn,             
                   Bonn, Germany}                                             
\centerline{$^{22}$Physikalisches Institut, Universit{\"a}t Freiburg,         
                   Freiburg, Germany}                                         
\centerline{$^{23}$Institut f{\"u}r Physik, Universit{\"a}t Mainz,            
                   Mainz, Germany}                                            
\centerline{$^{24}$Ludwig-Maximilians-Universit{\"a}t M{\"u}nchen,            
                   M{\"u}nchen, Germany}                                      
\centerline{$^{25}$Fachbereich Physik, University of Wuppertal,               
                   Wuppertal, Germany}                                        
\centerline{$^{26}$Panjab University, Chandigarh, India}                      
\centerline{$^{27}$Delhi University, Delhi, India}                            
\centerline{$^{28}$Tata Institute of Fundamental Research, Mumbai, India}     
\centerline{$^{29}$University College Dublin, Dublin, Ireland}                
\centerline{$^{30}$Korea Detector Laboratory, Korea University,               
                   Seoul, Korea}                                              
\centerline{$^{31}$SungKyunKwan University, Suwon, Korea}                     
\centerline{$^{32}$CINVESTAV, Mexico City, Mexico}                            
\centerline{$^{33}$FOM-Institute NIKHEF and University of                     
                   Amsterdam/NIKHEF, Amsterdam, The Netherlands}              
\centerline{$^{34}$Radboud University Nijmegen/NIKHEF, Nijmegen, The          
                  Netherlands}                                                
\centerline{$^{35}$Joint Institute for Nuclear Research, Dubna, Russia}       
\centerline{$^{36}$Institute for Theoretical and Experimental Physics,        
                   Moscow, Russia}                                            
\centerline{$^{37}$Moscow State University, Moscow, Russia}                   
\centerline{$^{38}$Institute for High Energy Physics, Protvino, Russia}       
\centerline{$^{39}$Petersburg Nuclear Physics Institute,                      
                   St. Petersburg, Russia}                                    
\centerline{$^{40}$Lund University, Lund, Sweden, Royal Institute of          
                   Technology and Stockholm University, Stockholm,            
                   Sweden, and}                                               
\centerline{Uppsala University, Uppsala, Sweden}                              
\centerline{$^{41}$Physik Institut der Universit{\"a}t Z{\"u}rich,            
                   Z{\"u}rich, Switzerland}                                   
\centerline{$^{42}$Lancaster University, Lancaster, United Kingdom}           
\centerline{$^{43}$Imperial College, London, United Kingdom}                  
\centerline{$^{44}$University of Manchester, Manchester, United Kingdom}      
\centerline{$^{45}$University of Arizona, Tucson, Arizona 85721, USA}         
\centerline{$^{46}$Lawrence Berkeley National Laboratory and University of    
                   California, Berkeley, California 94720, USA}               
\centerline{$^{47}$California State University, Fresno, California 93740, USA}
\centerline{$^{48}$University of California, Riverside, California 92521, USA}
\centerline{$^{49}$Florida State University, Tallahassee, Florida 32306, USA} 
\centerline{$^{50}$Fermi National Accelerator Laboratory,                     
            Batavia, Illinois 60510, USA}                                     
\centerline{$^{51}$University of Illinois at Chicago,                         
            Chicago, Illinois 60607, USA}                                     
\centerline{$^{52}$Northern Illinois University, DeKalb, Illinois 60115, USA} 
\centerline{$^{53}$Northwestern University, Evanston, Illinois 60208, USA}    
\centerline{$^{54}$Indiana University, Bloomington, Indiana 47405, USA}       
\centerline{$^{55}$University of Notre Dame, Notre Dame, Indiana 46556, USA}  
\centerline{$^{56}$Purdue University Calumet, Hammond, Indiana 46323, USA}    
\centerline{$^{57}$Iowa State University, Ames, Iowa 50011, USA}              
\centerline{$^{58}$University of Kansas, Lawrence, Kansas 66045, USA}         
\centerline{$^{59}$Kansas State University, Manhattan, Kansas 66506, USA}     
\centerline{$^{60}$Louisiana Tech University, Ruston, Louisiana 71272, USA}   
\centerline{$^{61}$University of Maryland, College Park, Maryland 20742, USA} 
\centerline{$^{62}$Boston University, Boston, Massachusetts 02215, USA}       
\centerline{$^{63}$Northeastern University, Boston, Massachusetts 02115, USA} 
\centerline{$^{64}$University of Michigan, Ann Arbor, Michigan 48109, USA}    
\centerline{$^{65}$Michigan State University,                                 
            East Lansing, Michigan 48824, USA}                                
\centerline{$^{66}$University of Mississippi,                                 
            University, Mississippi 38677, USA}                               
\centerline{$^{67}$University of Nebraska, Lincoln, Nebraska 68588, USA}      
\centerline{$^{68}$Princeton University, Princeton, New Jersey 08544, USA}    
\centerline{$^{69}$State University of New York, Buffalo, New York 14260, USA}
\centerline{$^{70}$Columbia University, New York, New York 10027, USA}        
\centerline{$^{71}$University of Rochester, Rochester, New York 14627, USA}   
\centerline{$^{72}$State University of New York,                              
            Stony Brook, New York 11794, USA}                                 
\centerline{$^{73}$Brookhaven National Laboratory, Upton, New York 11973, USA}
\centerline{$^{74}$Langston University, Langston, Oklahoma 73050, USA}        
\centerline{$^{75}$University of Oklahoma, Norman, Oklahoma 73019, USA}       
\centerline{$^{76}$Oklahoma State University, Stillwater, Oklahoma 74078, USA}
\centerline{$^{77}$Brown University, Providence, Rhode Island 02912, USA}     
\centerline{$^{78}$University of Texas, Arlington, Texas 76019, USA}          
\centerline{$^{79}$Southern Methodist University, Dallas, Texas 75275, USA}   
\centerline{$^{80}$Rice University, Houston, Texas 77005, USA}                
\centerline{$^{81}$University of Virginia, Charlottesville,                   
            Virginia 22901, USA}                                              
\centerline{$^{82}$University of Washington, Seattle, Washington 98195, USA}  
}                                                                             

\affiliation{}

\collaboration{D\O\ Collaboration}

\date{September 19, 2006}

\begin{abstract}
We report on a measurement of the $B^0_d$ mixing frequency and the
calibration of an opposite-side flavor tagger in the D\O\ experiment.
Various properties associated with the $b$ quark on the opposite side
of the reconstructed $B$ meson were combined using a likelihood-ratio method
into a single variable with enhanced tagging power. Its performance was 
tested with data, using a large sample of reconstructed  semileptonic 
$B \to \mu \dzero X$ and $B \to \mu \dst X$ decays, corresponding to 
an integrated luminosity of approximately 1 fb$^{-1}$.
The events were divided into groups depending on the value of the combined 
tagging variable, and an independent analysis was performed in each group. 
Combining the results of these analyses, the overall effective tagging power 
was found to be 
$\merit = (2.48 \pm 0.21 ^{+0.08}_{-0.06})\%$. 
The measured $B^0_d$ mixing frequency 
$\dmd = 0.506 \pm 0.020~{\rm (stat)~ \pm 0.016~ (syst) ~ps}^{-1}$
is in good agreement with the world average value.
\end{abstract}

\pacs{}

\maketitle

\section{Introduction \label{intro}}
Particle-antiparticle mixing in the $B^{0}$ ($B^0_d$) system has 
been known for more than a decade now \cite{bdmix} and has been
studied at the CERN LEP collider and subsequently at the Fermilab 
Tevatron collider during Run I. It is currently being measured at the 
$B$-factory experiments, Belle and BaBar, 
and the Fermilab Tevatron collider experiments during Run II.

Mixing measurements involve identifying the ``flavor'' of the $B^{0}$ meson
at production and again when it decays, where ``flavor'' indicates whether
the meson contained a $b$ or a $\bar{b}$ quark. The decay flavor is identified
from the $B^0$ decay products when the $B^{0}$ meson is reconstructed. The 
determination of the initial ``flavor'' is known as flavor tagging.

\begin{figure}[htb]
    \includegraphics[width=9.0cm]{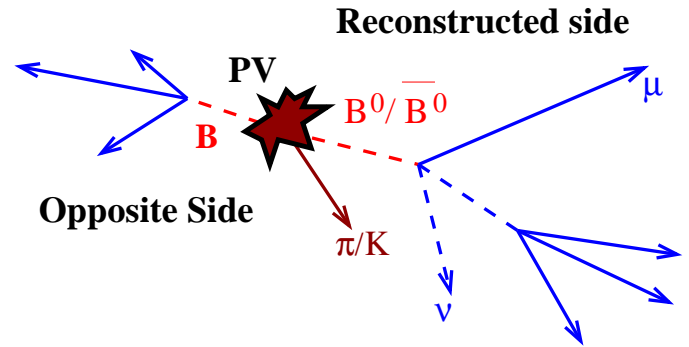}
    \caption[]{
        \label{intro:fig1} Diagram of an event with a reconstructed $B^0$ candidate. PV indicates the primary vertex for the event. 
    }
\end{figure}

The $B^0_d$ meson flavor at its production can be identified using information
from the reconstructed side or from the opposite side 
(see Fig.~\ref{intro:fig1}). One can tag the flavor using charge 
correlation between ``fragmentation tracks'' associated with the 
reconstructed $B$ meson. Such correlations were first observed in 
$e^{+} e^{-} \rightarrow Z^{0} \rightarrow b \bar{b}$ events by the
OPAL experiment
\cite{sst}. This is known as ``same-side flavor tagging.'' 
The flavor can also be inferred from the decay information of the second 
$B$ meson in the event, assuming that $b$ and $\bar{b}$ are produced in pairs, 
and thus in the ideal case, the two mesons have opposite flavors. This 
method is known as ``opposite-side flavor tagging.'' An advantage of the latter 
method is that its performance should be independent of the type of the 
reconstructed $B$ meson.

Measurement of the $B^0_d$ mixing parameter is an important test of the 
opposite-side flavor tagging as the same tagger is used for our study of 
$B_s$ mixing. Studies of tagged $B^0$ and $B^{+}$ samples at hadron colliders 
could reveal physics beyond the standard model ~\cite{berger}. 
Finally, this technique of flavor tagging developed at the Tevatron can also 
be useful for future experiments at the Large Hadron Collider at CERN.

This paper describes the opposite-side flavor tagging algorithm used by the 
D\O\ experiment in Run II and the measurement of  its performance using 
$B \to \mu^+ \adzero X$ and $B \to \mu^+ \dstminus X$ events. 
Throughout the paper,a reference to a particular final state also implies its 
charge conjugated state. $B^+$ decays represent the main contribution to the 
$B \to \mu^+ \adzero X$ sample, and $B^0$ decays dominate in the
$B \to \mu^+ \dstminus X$ sample. 
We measure the flavor tagging purity independently for reconstructed 
$B^+$ and $B^0$ events and then extract the $B^0$ oscillation frequency. 
This technique allows us to verify the assumption of independence of the 
opposite-side flavor tagging on the type of reconstructed $B$ meson. Its 
performance is described by the two parameters, efficiency and dilution. 
The efficiency $\varepsilon$ is defined as the fraction of reconstructed 
events ($N_{\text {tot}}$) that are tagged ($N_{\text {tag}}$):

\begin{equation}
\varepsilon = N_{\text{tag}}/N_{\text{tot}}.
\end{equation} 

The dilution $\cal{D}$ is a normalized difference of correctly and wrongly tagged events: 
\begin{equation}\label{dil}
{\cal{D}} = \frac{N_{\text{cor}}-N_{\text {wr}}}{N_{\text {cor}}+N_{\text {wr}}}=\frac{N_{\text {cor}}-N_{\text {wr}}}{N_{\text {tag}}}=2P-1,
\end{equation} 

\noindent where $P=N_{\text {cor}}/N_{\text {tag}}$ is called the {\it purity}. The 
terms ``correctly'' and ``wrongly'' refer to the determination of the 
reconstructed $B$ meson flavor. The effective tagging power of a 
tagging algorithm is given by $\varepsilon {\cal{D}}^{2}$.

\section{\label{sec:det} Detector Description}
The D\O\ detector is described in detail elsewhere \cite{dzero}.
The main features of the detector essential for this analysis are summarized below.
Tracks of charged particles 
are reconstructed from the hits in the central tracking system, which consists of the 
silicon microstrip tracker (SMT) and the central fiber tracker
(CFT), both located in a $2$~T superconducting solenoidal magnet.
The SMT has $\approx 800,000$ individual strips, with a typical pitch of 
$50-80$~$\mu$m and a design optimized for tracking and vertexing capability for 
$|\eta|=3$. The pseudorapidity, $\eta = - \ln ~[\tan(\theta/2)]$, approximates the 
true rapidity, $y = \frac{1}{2} ~\ln [(E+p_{z}c)/(E-p_{z}c)]$, for finite angles in 
the limit of $({mc}^{2}/E) \rightarrow 0$, $\theta$ being the polar angle.
We use the term ``forward'' to describe 
the regions at large $\mid \eta \mid$. The SMT system consists of six
barrels arranged longitudinally (each with a set of four layers of silicon
detectors arranged axially around the beam pipe), interspersed with 16 radial
disks. The CFT has eight thin coaxial barrels, each supporting two doublets of 
overlapping scintillating fibers of 0.835~mm in diameter, one doublet being parallel 
to the beam axis and the other alternating by $\pm 3^{\circ}$ relative to 
this axis. Light signals are transferred via clear light fibers to solid-state 
photon counters (VLPCs) that have $\approx 80\%$ quantum efficiency.

The muon system consists of a layer of tracking detectors 
and scintillation trigger counters in front of  $1.8$~T toroids, 
followed by two additional similar layers after the toroids. Muon tracking for 
$|\eta|<1$ relies on $10$-cm-wide drift tubes, 
while $1$-cm mini drift tubes are used for $1<|\eta|<2$.

Electrons are identified using matching between the tracks
identified in the central tracker and energy deposits in a
primarily liquid-argon/uranium sampling calorimeter ~\cite{dzero}. 
We also use the energy deposits in the central preshower detector~\cite{dzero},
which consists of three concentric cylindrical layers of triangular
scintillator strips and is located in a nominal 5 cm gap
between the solenoid and the central calorimeter, to provide
additional discrimination between electrons and fakes.
The calorimeter consists of the inner electromagnetic section 
followed by the fine and coarse hadronic sections. In this analysis, we
only use the central calorimeter ($\mid \eta \mid < 1$). 
 
\section{Data Sample and Event Selection}
\label{sec:measmeth}
This measurement is based on a large semileptonic $B$ decay data 
sample corresponding to approximately 1~fb$^{-1}$ of integrated
luminosity collected with the D\O\ detector between April 2002 and October 
2005. 

\label{sec:evsel}
$B$ mesons were selected using their semileptonic decays
$B \rightarrow \mu^+\nu\bar{D}^0X$ 
and were divided into two exclusive groups: the \dst sample, 
containing all events with reconstructed $D^{*-}\to\bar{D}^0\pi^-$ decays, 
and the \dzero sample, containing all the remaining events.
The \dst sample is dominated by $B^0_d \to \mu^+ \nu \dstminus X$ decays, 
while the \dzero sample is dominated by $B^+ \to \mu^+ \nu \adzero X$ decays.

The flavor tagging procedure was developed using events from the \dzero sample.
Events from the \dst sample were used to measure the purity of the flavor tagging 
and the oscillation
parameter \dmd. In addition, the purity was measured in the \dzero
sample to test the hypothesis that the flavor tagger is independent of the type
of reconstructed $B$ meson.

Muons for this analysis were required to 
have hits in more than one muon chamber, an associated track in the 
central tracking system with hits in both SMT and CFT detectors, transverse 
momentum $p_T^{\mu} > 2$ GeV/$c$, as measured in the central tracker, 
pseudorapidity $|\eta^\mu|<2$, and total momentum $p^{\mu} > 3$ GeV/$c$.

All charged particles in a given event were clustered into jets using the
DURHAM clustering algorithm \cite{durham} with the cut-off parameter set to 15 GeV/$c$. Events with more than one identified muon in the same jet or with 
the reconstructed $J/\psi \to \mu^+ \mu^-$ decays were rejected.

\dzero candidates were constructed from two tracks of opposite charge 
belonging to the same jet as the reconstructed muon. 
Both tracks were required to have transverse momentum $p_T > 0.7$ GeV/$c$ and 
pseudorapidity $|\eta|<2$. 
They were required to form a common $D$ vertex with a 
fit $\chi^2 < 9$, number of degrees of freedom being 1. 
For each track, the projection $\epsilon_T$ 
(onto the axial plane, i.e. perpendicular to the beam direction) and projection
$\epsilon_L$ (onto the stereo plane, i.e. parallel to the beam direction) 
of its impact parameter with respect to the 
primary vertex, together with the corresponding uncertainties ($\sigma(\epsilon_T)$, 
$\sigma(\epsilon_L)$) were computed.
The combined impact parameter significance 
$S = \sqrt{[\epsilon_T/\sigma(\epsilon_T)]^2 + [\epsilon_L/\sigma(\epsilon_L)]^2}$ 
was required to be greater than 2.
The distance $d_T^D$ between the primary and $D$ vertices in the axial plane 
was required to exceed 4 standard deviations: $d_T^D/\sigma(d_T^D) > 4$. 
The accuracy of the $d_T^D$ determination was required to be better 
than $500$~$\mu$m. The angle $\alpha^D_T$ between the \dzero momentum vector 
and 
the direction from the primary to the $D$ vertex in the axial plane was 
required to satisfy the condition $\cos \alpha^D_T > 0.9$. The tracks of the 
muon and \dzero candidate were required to form a common $B$ vertex
with a fit $\chi^2 < 9$, with number of degrees of freedom being 1. 
The mass of the kaon was assigned to the track having the same charge as
the muon; the remaining track was assigned the mass of the pion.
The mass of the $(\mu^+ \adzero)$ 
system was required to fall within the $2.3 < M(\mu^+ \adzero) < 5.2$ GeV/$c^2$
range.

If the distance $d_T^B$ between the primary and $B$ vertices in the
 axial plane exceeded $4\sigma(d_T^B)$, the angle $\alpha^B_T$ between 
the $B$ momentum and the direction from the primary to the $B$ vertex 
in the axial plane was required 
to satisfy the condition $\cos \alpha^B_T > 0.95$. The distance $d_T^B$ was 
allowed to be greater than $d_T^D$, provided that the distance between the
$B$ and $D$ vertices $d_T^{BD}$ was less than $3\sigma(d_T^{BD})$.
The uncertainty $\sigma(d_T^B)$ was required to be less than $500$~$\mu$m.
In addition, the cut $p_T(\adzero) > 5$ GeV/$c^2$ was applied. 

\begin{figure}
\begin{center}
\includegraphics[width=8.0cm]{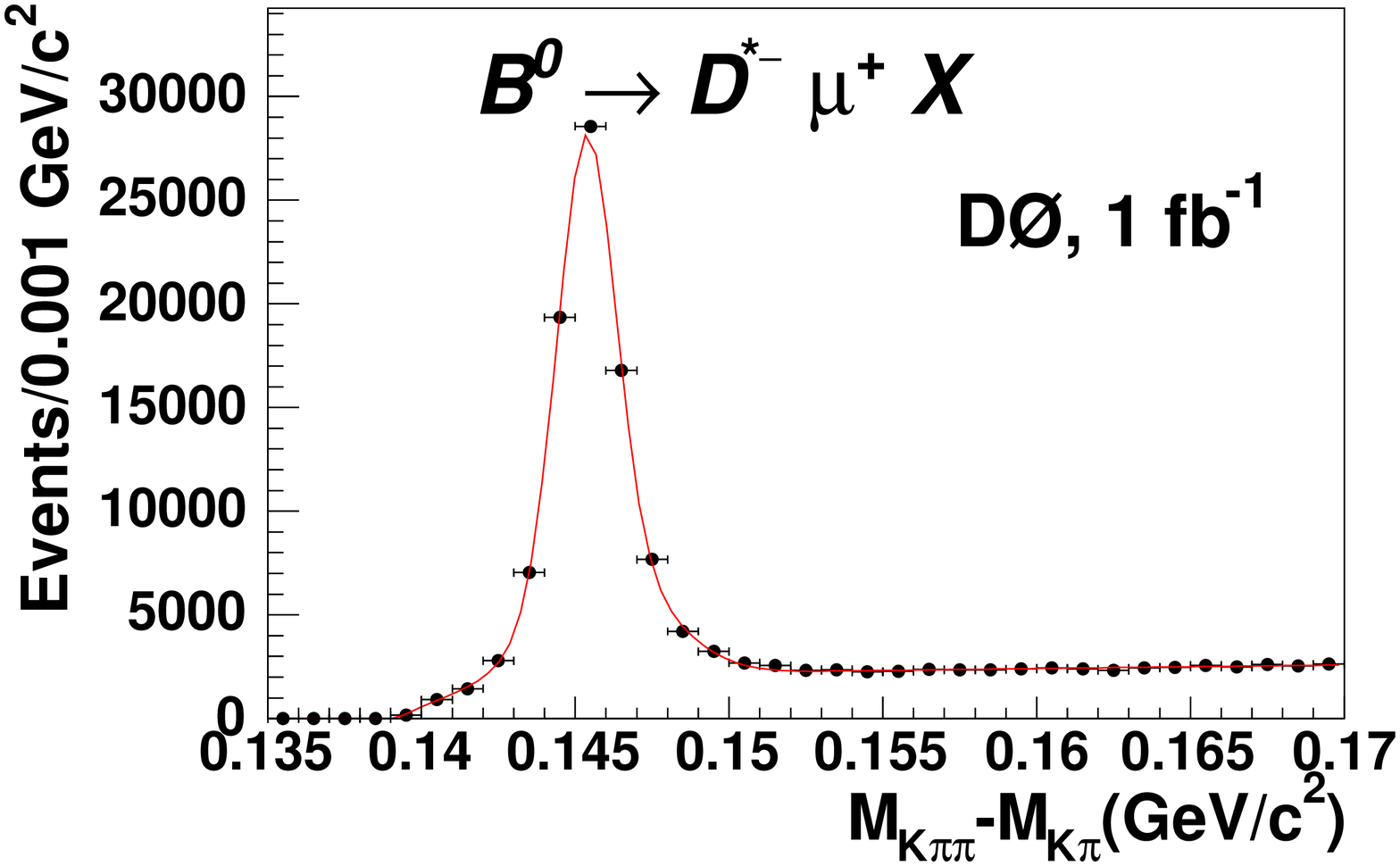}
\end{center}
\caption{\label{fig:fig3}
  The $M(K \pi \pi) - M(K \pi)$ invariant mass distribution for selected $\mu \dst$ candidates. 
  The curve shows the result of the fit described in Sec.\ref{sec:massfitproc}.
}
\end{figure}

To select $\mu^+ \dstminus$ candidates, we searched for an additional pion track
with $p_T > 0.18$~GeV/$c$ and the charge opposite to the charge of the muon.
The mass difference $\Delta M = M(K \pi \pi) - M(K \pi)$ for \dst
candidates, with  $1.75 < M(\adzero) < 1.95$ GeV/$c^2$, is shown
in Fig.~\ref{fig:fig3}. The peak corresponding to the mass of the
soft pion in the  $\mu^+ \dstminus$ sample is clearly seen.  

\begin{figure}
\begin{center}
\includegraphics[width=8.5cm]{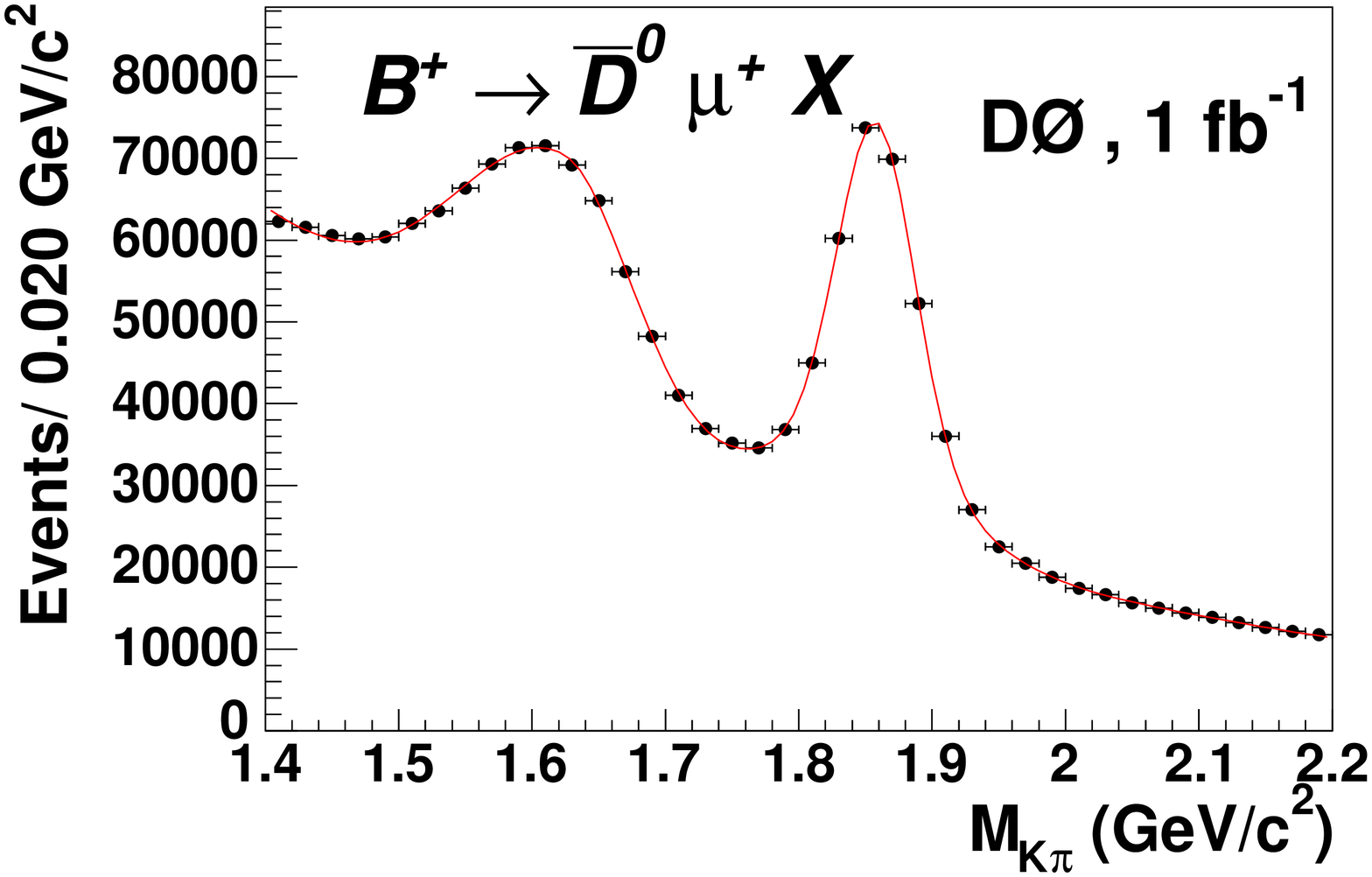} \\
\caption{\label{fig:fig1}
   The $K \pi$ invariant mass distribution for selected $\mu \dzero$ candidates. 
   The curve shows the result of the fit 
   described in Sec.\ref{sec:massfitproc}.
}
\end{center}
\end{figure}

\begin{figure}
\begin{center}
\includegraphics[width=8.5cm]{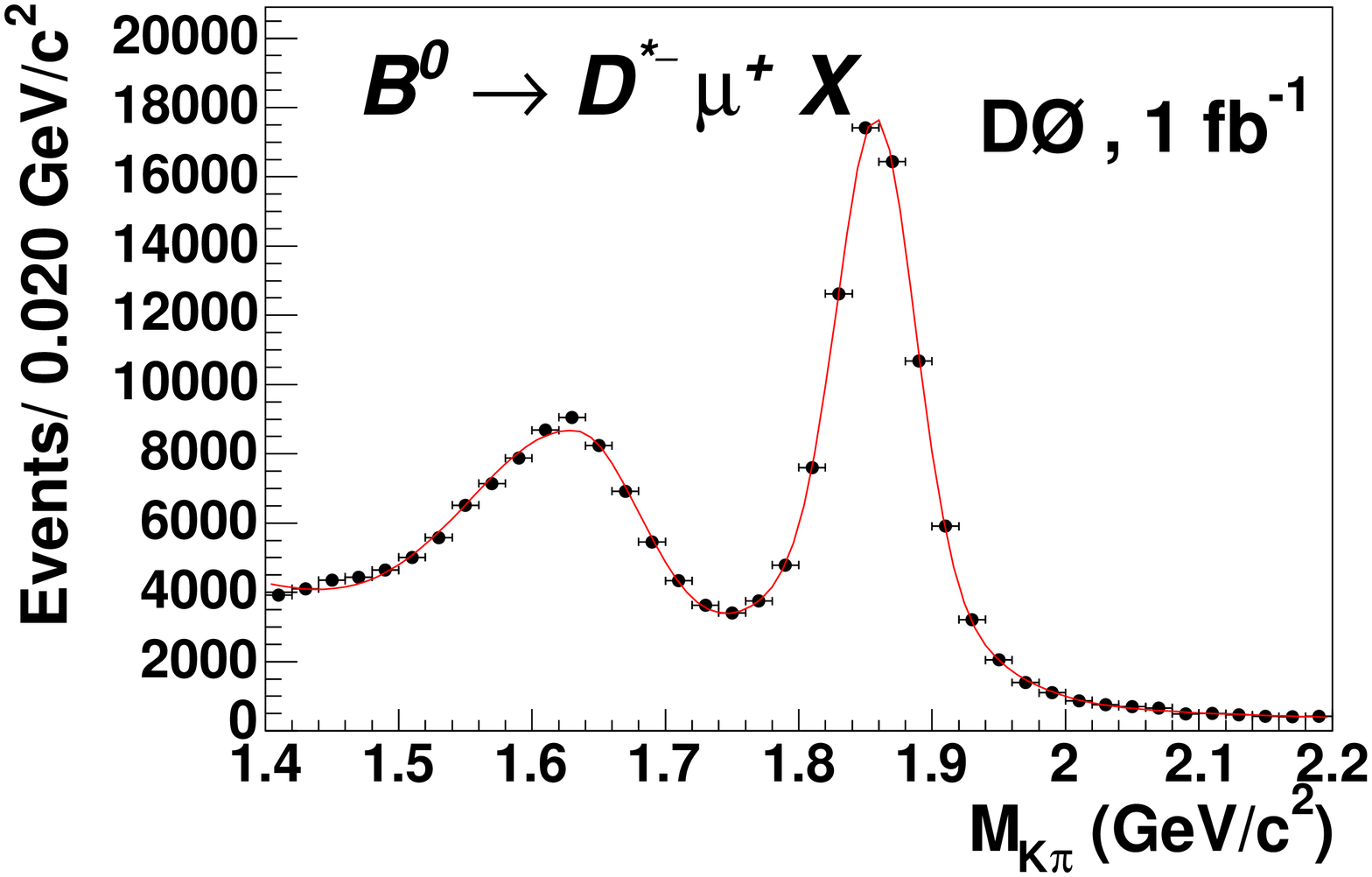}
\end{center}
\caption{\label{fig:fig2}
  The $K \pi$ invariant mass for selected $\mu \dst$ candidates. 
  The curve shows the result of the fit described in Sec.\ref{sec:massfitproc}.
}
\end{figure}

All events with $0.1425 < \Delta M < 0.1490$ GeV/$c^2$ were included in the \dst
sample. The remaining events were assigned to the \dzero sample. The $K \pi$
mass distributions for these two samples together with the results of the fits
are shown in Figs. \ref{fig:fig1} and \ref{fig:fig2}. The procedure to fit
these mass spectra is described in Sec.\ref{sec:massfitproc}.
In total, $230551 \pm 1627$ $B \to \mu^+ \nu \adzero$ decays 
and $73532 \pm 304$ $B \to \mu^+ \nu \adst$ decays 
were reconstructed.

\section{Visible Proper Decay Length}

The oscillations of $B$ mesons are usually 
studied as a function of their proper decay 
length. Since in semileptonic $B$ decays an undetected neutrino 
carries away part of the energy, the proper decay length cannot be accurately measured. 
Instead, a {\it visible proper decay length} (VPDL) was used in this analysis.
It is defined as

\begin{equation}
\label{vpdl}
L = M_B (\bm{L_{xy}}\cdot\bm{P_{xy}^{\mu \dzero}})/
|\bm{P_{T}^{\mu \dzero}}|^2.
\end{equation}

Here $\bm{L_{xy}}$ is a
vector in the plane perpendicular to the beam direction
from the primary to the $B$ meson decay vertex.
The transverse momentum $\bm{P_T^{\mu \dzero}}$ is defined
as the vector sum of the transverse momenta of the muon and $D^{0}$. $M_B$ 
is the mass of $B$ meson. 

\section{Description and combination of flavor taggers}

\label{sec:ist}

Many different properties can be used to identify the initial flavor 
$b$ or $\bar{b}$ of a heavy quark fragmenting into a reconstructed $B$ meson. 
Some of them have strong separation power, while others are
weaker. In all cases, their combination into a single tagging
variable gives a significantly better result than that in the case of
their separate use. We build such a combination
with the likelihood ratio method described below. We first describe
the combination algorithm and then discuss the discriminating
variables used.

\subsection{Combination of variables}
We construct a set of discriminating variables $x_1, ..., x_n$ for a given event. 
A discriminating variable,
by definition, should have different distributions for $b$ and $\bar{b}$ flavors. 
For the initial $b$ quark,
the probability density function (p.d.f.) for a given variable $x_i$
is denoted as $f_i^b(x_i)$, while for the initial $\bar{b}$ quark it is
denoted as $f_i^{\bar{b}}(x_i)$. The combined tagging variable $r$
is defined as: 
\begin{equation}
\label{ctag}
r = \prod_{i=1}^{n}r_i;~~~ r_i = \frac{f_i^{\bar b}(x_i)}{f_i^{b}(x_i)}.
\end{equation}
A given variable $x_i$ may not be defined for some events. For example, there are
events that do not contain an identified muon on the opposite side.
In this case, the corresponding variable $r_i$ is set to 1. 
An initial $b$ flavor is more probable if $r<1$, and a
$\bar{b}$ flavor is more probable if $r>1$. By construction, an event
with $r<1$ is tagged as a $b$ quark and an event with $r>1$ is tagged as
a $\bar{b}$ quark.
For an oscillation analysis, it is more convenient to define the tagging
variable as:
\begin{equation}
\label{ctag1}
 d = \frac{1-r}{1+r},
\end{equation}
By construction, the variable $d$ which ranges between $-1$ and $1$.
An event with $d>0$ is tagged as a $b$ quark and with $d<0$ as a $\bar{b}$
quark, with higher $|d|$ values corresponding to higher tagging purities.
For uncorrelated variables $x_1, ..., x_n$, and perfect modeling
of the {p.d.f.}, $d$ gives the best
possible tagging performance, and its absolute value provides a measure of
the dilution of the flavor tagging defined in Eqn. \ref{dil}.

Very often, the analyzed events are divided into samples
with significantly different discriminating variables and 
tagging performances.
This division would imply making a separate analysis for each sample
and combining the results at a later stage.
In contrast to this approach, the tagging variable $d$ defined by Eqs. 
\ref{ctag} and \ref{ctag1} 
provides a ``calibration'' for all events, regardless of their intrinsic
differences. Since the absolute value of $d$ gives a measure of
the dilution of the flavor tagging, events from different categories 
but with a similar absolute value of $d$ can be treated in the same way.
Thus, another important advantage of this  method of flavor
tagging is the possibility of building a single variable having the same
meaning for different kinds of events. It allows us to 
classify all events according to their tagging characteristics
and use them simultaneously in the analysis.

\label{sec:discs}

All of the discriminating variables used in this analysis 
are constructed using the properties of the $b$ quark opposite to the reconstructed $B$ meson
(``opposite-side tagging'').
Since an important property of the opposite-side tagging is the independence
of its performance of the type of reconstructed $B$ meson, 
it can be calibrated 
in data by applying tagging to the events with $B^0$ and $B^+$ decays.
The measured performance can then be used to study $B^0_s$ meson oscillations, 
as an example.

The probability density functions for each discriminating variable
discussed below were constructed using events from the \dzero sample with
$0 < {\text {VPDL}} < 500 ~\mu$m. 
In this sample, the decay $B^+ \to \mu^+ \nu \adzero$ 
dominates, see Sec. \ref{sec:istbd}.
The $B^0_d \to \mu^+ \nu \dstplus$ events give a ~16\% contribution
to the sample and, due to the cut on VPDL, contains mainly non-oscillated $B^0_d$ decays, as determined by Monte Carlo (the standard pythia
generation, followed by decay of B mesons with EvtGen, passed through
Geant and then reconstruction).

The initial flavor of a $b$ quark is therefore determined by the charge 
of the muon. Estimates based on Monte Carlo simulation indicate that
the purity of the initial flavor determination in the selected sample is 
$0.98 \pm 0.01$, where the uncertainty is due to the uncertainties in 
measured branching fractions of $B$ meson decays. 

For each discriminating variable,
the signal band containing all events
with $1.80 < M(K \pi) < 1.92$ \GVcs~ and the background band
containing all events with $1.94 < M(K \pi) < 2.2$ \GVcs~ were defined.
The p.d.f's were constructed as the difference in the distributions. The 
latter distributions were normalized by multiplying them by 0.74
so that the number of events in the background band corresponds to the 
estimated number of background events in the signal band.

\subsection{Flavor tagger discriminants}\label{subsec:discr}
\begin{figure}[htb]
    \begin{center}
    \includegraphics[width=9.0cm]{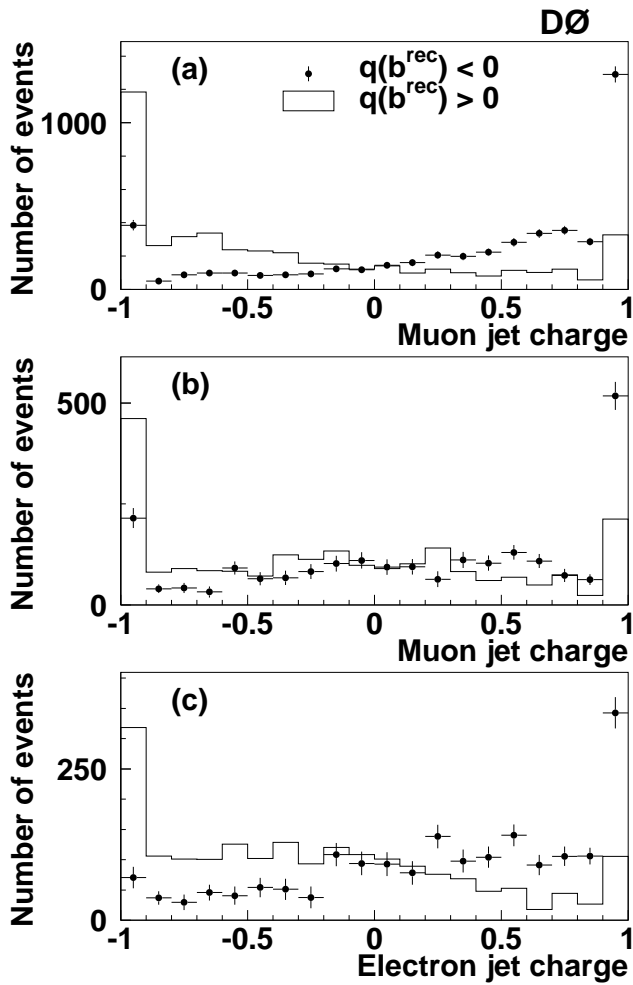}
    \caption[]{
        \label{ftag:fig1} 
	(a) Distribution of the jet charge $Q_J^{\mu}$ for muons with hits in all 
	three layers of the muon detector.
	(b) Distribution of the jet charge $Q_J^{\mu}$, for muons with fewer than three hits.
		(c) Distribution of the jet charge for electrons $Q_J^{e}$. Here
        $q(b^{\text {rec}})$ is the
        charge of the muon from the reconstruction side.
    }	
   \end{center}
\end{figure}

We now describe the variables used.
An additional muon was searched for in each analyzed event. This muon
was required
to have at least one hit in the muon chambers
and to have $\cos \phi({\bm p}_\mu,{\bm p}_B) < 0.8$, 
where ${\bm p}_B$ is the three-momentum of the reconstructed $B$ meson, 
and $\phi$ is the angle between the vectors ${\bm p}_\mu$ and
${\bm p}_B$.
If more than one muon was found, the muon with the
highest number of hits in the muon chambers was used. If more than
one muon with the same number of hits in the muon chambers was found,
the muon with the highest transverse momentum $p_T$ was used.
For this muon, a {\it muon jet charge} $Q_J^\mu$ was constructed as
\[ Q_J^\mu = \frac{\sum_i {q^i p_T^i}}{\sum_i {p^i_T}}, \]
where $q^i$ is the charge and $p_T^i$ is the transverse momentum 
of the $i$'th particle, and 
the sum is taken over all charged particles, including the muon,
satisfying the condition 
$\Delta R = \sqrt{(\Delta \phi)^2 + (\Delta \eta)^2} < 0.5$,
where $\Delta \phi$ and $\Delta \eta$ are computed with respect to 
the muon direction.  Daughters of the reconstructed $B$ meson
were explicitly excluded from this sum. In addition, any charged particle
with $\cos \phi({\bm p},{\bm p}_B) > 0.8$ was excluded.
The distribution of the muon jet charge variable 
is shown in Figs. \ref{ftag:fig1}(a) and (b).
In these plots, $q^{\text {rec}}$ gives 
the charge of the $b$ quark in the reconstructed 
$B \to \mu^+ \nu \adzero$ decay, in this case given
by the muon charge.
We build separate p.d.f.'s for muons with hits in all 
three layers of the muon detector, Fig.~\ref{ftag:fig1}(a), and for 
muons with fewer than three hits,  Fig.\ \ref{ftag:fig1}(b).

In addition to the muon tag, reconstructed electrons with $\cos \phi({\bm p}_e,{\bm p}_B) < 0.8$ were also used for flavor tagging. 
The electron is reconstructed by extrapolating a track to the calorimeter and adding up the energy deposited in a narrow tube or ``road'' around the track. Calorimeter cells
are collected around the track extrapolated positions in each layer
and the total transverse energy of the cluster is defined by the sum of the 
energies in each layer. The electrons are required to be in the central region 
($|\eta| < 1.1$), with  $p_T > 2$ GeV/c. They are required to have at 
least one hit each in the CFT and SMT. They are required to have energy deposits 
in the EM calorimeter consistent with an electron, 
$ 0.55(0.5) < E/p < 1.0(1.1)$, and low energy deposit in the hadron calorimeter, $ EMF > 0.8(0.7)$. The cuts are looser for electrons with $p_T > 3.5$ \GVc~ and are given
in brackets. $EMF$ and $E/p$ are calculated as below:
\begin{eqnarray}
EMF & = & \frac{\sum_{\rm layer ~number ~i=1,2,3} E_{T}(i)}{\sum_{\rm all layers} E_{T} (i)} \\
E/p & = & \frac{\sum_{\rm layer ~number ~i=1,2,3} E_{T}(i)}{p_{T} (track)},
\end{eqnarray}
where $E_T(i)$ is the transverse energy within the road in the $i$'th layer.
We also require a minimum single layer cluster energy of a cluster 
in the central preshower, ${\rm CPS}^{\rm SLC}_E > 4.0 (2.0)$ MeV/c.
The cuts were optimized by studying electrons from conversion decays ($\gamma \rightarrow e^{+} e^{-}$) and fakes from $K^0_{S} \rightarrow \pi^{+} \pi^{-}$ decays to obtain a $90 \%$ purity for electrons.
For these electrons, an {\it electron jet charge} ($Q_J^e$) was constructed in
the same way as the {\it muon jet charge}, $Q_J^{\mu}$. 
The distribution of the electron jet charge variable 
is shown in Fig.\ \ref{ftag:fig1}(c).

An additional secondary vertex corresponding to the decay of a $B$ hadron
was searched for, using all charged tracks in the event
excluding those from the reconstructed $B$ hadron.
The secondary vertex was also required to contain at least two tracks with an
axial impact parameter significance greater than 3. The distance $l_{xy}$ from 
the primary to the secondary vertex must also satisfy the condition:
$l_{xy} > 4 \sigma(l_{xy})$. The details of the secondary vertex
identification algorithm can be found in Ref. \cite{pvreco}. 

\begin{figure}[tbh]
    \begin{center}
    \includegraphics[width=9.0cm]{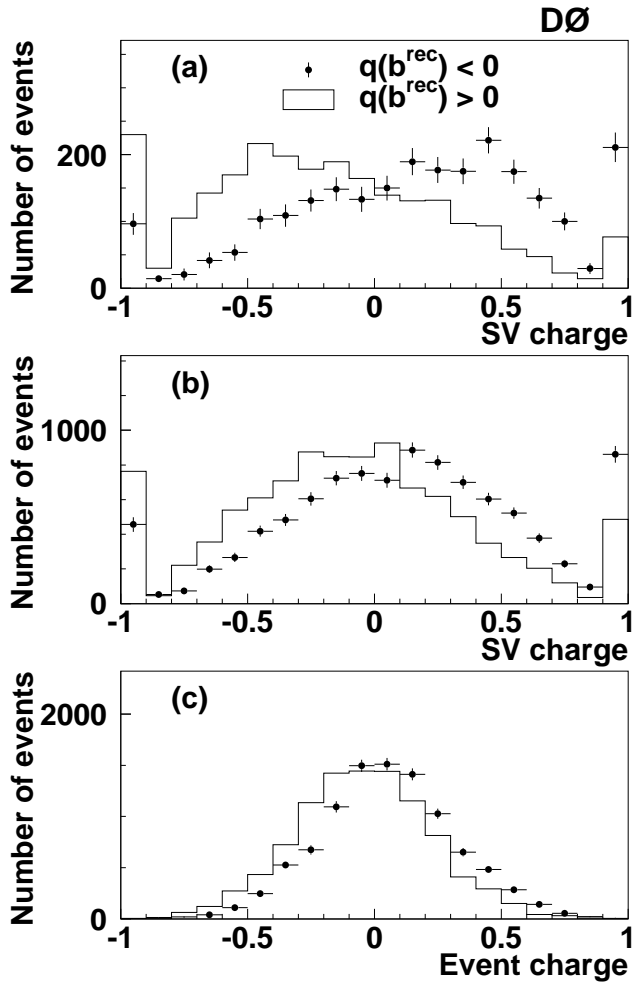}
    \caption[]{
    \label{ftag:fig2}  
	(a) Distribution of the secondary vertex charge for events with an opposite-side muon.
	(b) Distribution of the secondary vertex charge for events without an opposite-side muon.
	(c) Distribution of the event jet charge.
        $q(b^{\text {rec}})$ is the
        charge of the $b$ quark from the reconstruction side.
    }
  \end{center}
\end{figure}

The three-momentum of the secondary vertex ${\bm p}_{SV}$ is defined 
as the vector sum of the momenta of all tracks included in the secondary vertex. 
A secondary vertex with $\cos \phi({\bm p}_{SV},{\bm p}_B) < 0.8$ was
used for flavor tagging. 
A {\it secondary vertex charge} $Q_{SV}$ is defined as the third 
discriminating variable
\[ Q_{SV} = \frac{\sum_i {(q^i p_L^i)^k}}{\sum_i {(p^i_L)^k}}, \]
where the sum is taken over all tracks included in the secondary vertex.
Daughters of the reconstructed $B$ meson were explicitly excluded 
from this sum. In addition, any charged particle
with $\cos \phi({\bm p},{\bm p}_B) > 0.8$ was excluded.
Here $p_L^i$ is the longitudinal momentum of track $i$ with respect to 
the direction of the secondary vertex momentum {\bf p}$_{\sl V}$. 
A value of $k = 0.6$ was used, taken from previous studies
at LEP \cite{delphi}. We verified that this value of $k$ results in the optimal
performance of the $Q_{SV}$ variable.
Figures \ref{ftag:fig2}(a) and \ref{ftag:fig2}(b) show the distribution
of this variable for the events with and without an identified muon flavor tag.

Finally, the {\it event charge} $Q_{EV}$ was constructed as
\[ Q_{EV} = \frac{\sum_i {q^i p_T^i}}{\sum_i {p^i_T}}. \]
The sum is taken over all charged tracks with $0.5 < p_T < 50$ \GVc~
and having $\cos \phi({\bm p},{\bm p}_B) < 0.8$.
Daughters of the reconstructed $B$ meson were explicitly excluded 
from this sum. 
The distribution of this variable is shown in Fig.\ \ref{ftag:fig2}(c).

\begin{figure}[tbh]
    \includegraphics[width=9.0cm]{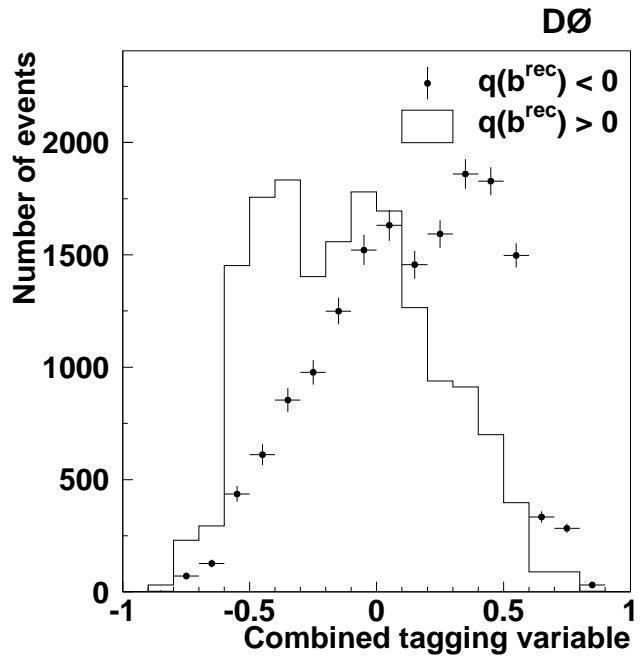}
    \caption[]{
    \label{ftag:fig3}  
	Normalized distributions of the combined tagging variable.
        $q(b^{\text rec})$ is the
        charge of the $b$ quark from the reconstruction side.
    }
\end{figure}

\label{sec:combined}
For each event with an identified muon, the muon jet charge $Q_J^\mu$
and the secondary vertex charge $Q_{SV}$ were used to construct 
a {\it muon tagger}. For each event without a muon but with an identified electron, 
the electron jet charge $Q_J^e$ and the secondary vertex charge $Q_{SV}$ were used to construct 
an {\it electron tagger}. Finally, for events without a muon or an electron but with a reconstructed
secondary vertex, the secondary vertex charge $Q_{SV}$
and the event jet charge $Q_{EV}$ were used to construct 
a {secondary vertex tagger}. The resulting distribution
of the tagging variable $d$ for the combination of all three taggers,
called the {combined tagger}, is shown in Fig. \ref{ftag:fig3}.
The performances of these taggers are discussed in the following sections.

\section{Multidimensional Tagger}
\label{sec:mdimtag}
In addition to the flavor tagger described in Sec. \ref{sec:ist}, 
an alternative 
algorithm was also developed and used to measure \bzerod~ mixing.  
This tagger
is \emph{multidimensional}, i.e., the likelihood functions it is
based on depend on more than a single variable.  
In addition, the  p.d.f.'s were determined from simulated events, 
while the primary flavor tagger described in Sec.\ref{sec:ist} 
uses data to construct the p.d.f.'s. The multidimensional tagger 
therefore provides a cross-check of the primary algorithm.

If, as before, we have a set of discriminants $x_1, ..., x_n$, 
the likelihood that the meson has flavor 
$b$ at the time of creation can be written as ${\cal L}(b;x_1, ..., x_n)$.  
A similar expression ${\cal L}(\bar{b};x_1, ..., x_n)$ holds
for the likelihood for $\bar{b}$.  These
likelihoods relate to the variable $d$ as
\begin{equation}
d=\frac{{\cal L}(b)-{\cal L}(\bar{b})}{{\cal L}(b)+{\cal L}(\bar{b})}.
\label{eq:d_likelihoods}
\end{equation}
This definition is similar to Eq. (\ref{ctag1}).

\label{sec:mdh_logic}
The likelihoods are obtained from the simulated samples 
of $B^{\pm} \rightarrow J/\psi K^{\pm}$ with $J/\psi \rightarrow 
\mu^{+}\mu^{-}$.   This final state does not oscillate and is therefore 
flavor-pure.  The $B^{-}\rightarrow J/\psi K^{-}$ sample was used
 to obtain ${\cal L}(b)$, while  ${\cal L}(\bar{b})$ was determined
from $B^{+}\rightarrow J/\psi K^{+}$ sample.
In practice, the likelihoods 
were stored as multidimensional histograms (with one dimension per discriminating 
variable) with the bin content normalized to the total number of events 
in the sample.  For a given event, the tagger output $d$ was obtained 
by substituting the appropriate normalized bin contents 
into Eq.\ (\ref{eq:d_likelihoods}).

In addition to the discriminating variables introduced in Sec.\ref{sec:ist},
other variables were used for the multidimensional tagger.
For each identified opposite-side muon, the 
transverse momentum $p_T$ relative to the beam axis and
transverse momentum $p_T^{\text {rel}}$ relative to the nearest jet were computed.
(The muon was included in the jet clustering.)
Another variable defined for the muon is its impact parameter significance 
$S_\mu$, where $S_\mu$ is the transverse impact parameter significance $\epsilon_{T}/\sigma(\epsilon_T)$, where $\epsilon_{T}$ is defined in Sec. \ref{sec:evsel}. 
For each reconstructed opposite-side secondary
vertex, the secondary vertex transverse momentum $p_{T}^{SV}$ was computed by
taking the magnitude of the transverse projection of the vector sum of all tracks 
in that vertex.
In principle, all discriminating variables
can be combined into a single multidimensional likelihood.
However, since a binned likelihood was used, in order to achieve a reasonable 
resolution in any given discriminant, the binning
must be fine enough to resolve its useful features.  
In practice, because of limited simulation statistics, this
means that discriminating variables must be chosen wisely 
when making a combination.

All events were divided into three categories based on their opposite-side 
content.  The following variables for different categories 
were selected.

\begin{enumerate}
\item Events with muon and secondary vertex: 
Tag($\mu+$SV)=$\left\{Q_{J}^{\mu};p_{T}^{\text{rel}};Q_{SV}\right\}$.
\item Events with muon and without secondary vertex.
Tag($\mu-$SV)=$\left\{Q_{J}^{\mu};p_{T}^{\text{rel}};p_{T};S_\mu \right\}$.
\item Events with secondary vertex without a muon: \\
Tag(SV$-\mu$)=$\left\{Q_{EV};Q_{SV};p_{T}^{SV}\right\}$.
\end{enumerate}

\begin{figure}[h]
    \begin{center}
    \includegraphics[width=10.0cm]{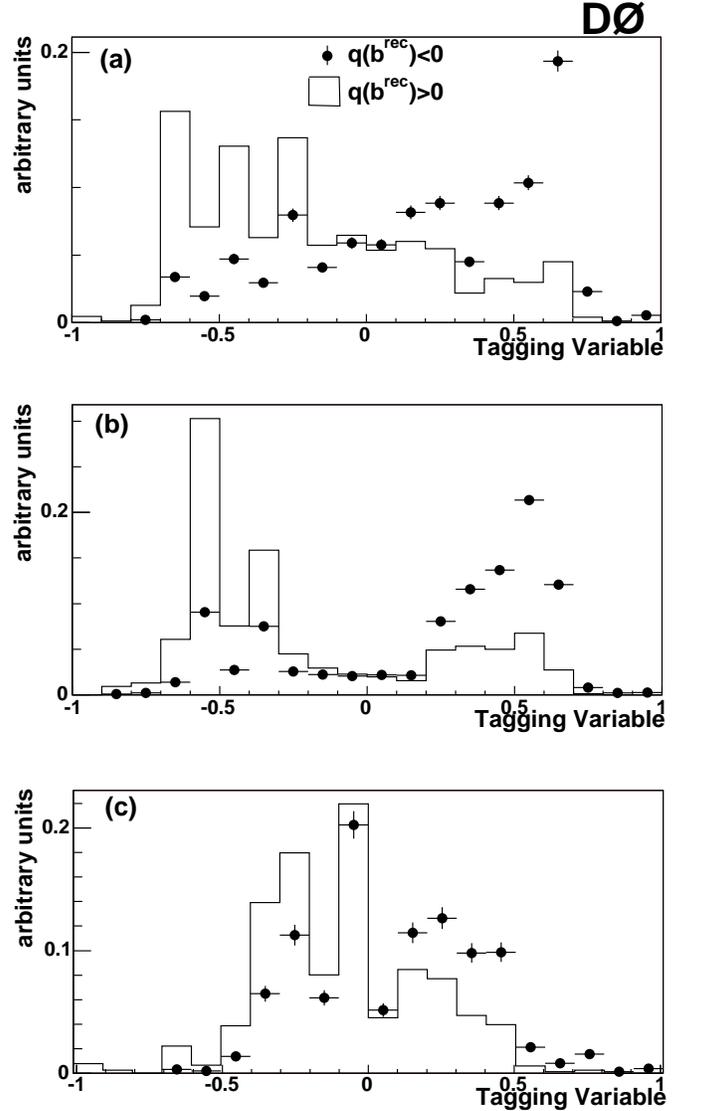}
    \caption[]{
        \label{fig:mdim_dvar} 
        Normalized distributions of the combined tagging variable 
for three multidimensional taggers for the simulated samples 
$B^{\pm}\rightarrow J/\psi K^{\pm}$.  
Here $q(b^{\text {rec}})$ is the charge of the $b$ quark from the reconstructed side.
	(a) Distribution of $d$ for Tag($\mu+$SV).
	(b) Distribution of $d$ for Tag($\mu-$SV).
	(c) Distribution of $d$ for Tag(SV$-\mu$).
     }	
   \end{center}
\end{figure}

Distributions in the tagging variable $d$ for the above three taggers 
are shown in Fig.\ \ref{fig:mdim_dvar}.  They were 
made by applying the taggers to the simulated 
$B^{\pm}\rightarrow J/\psi K^{\pm}$ samples from which they were created.

The final multidimensional tagger used the following logic to 
decide which of its sub-taggers to use.
For events containing a muon and a secondary vertex, the Tag($\mu+$SV)
was used. If the opposite side contained a muon and no secondary vertex, 
the Tag($\mu-$SV) was used.
If the opposite side contained an electron, the electron tagger 
described in Sec. \ref{sec:ist} was used.
Note that this tagger is not multidimensional 
and is not derived from simulation.
If the opposite side contained a secondary vertex, the Tag(SV$-\mu$)
was used.

\section{Asymmetry Fit Procedure}
\label{sec:massfitproc}

The performance of the flavor tagging and measurements of the \bzerod~ mixing
frequency \dmd~ were obtained from a study of the dependence of the flavor
asymmetry on the $B$-meson decay length.

The flavor asymmetry $A$ is defined as:
\begin{equation}
\label{asymobs}
A = \frac{N^{\text {nos}} - N^{\text {osc}}}{N^{\text{nos}} + N^{\text{osc}}}.
\end{equation}

\noindent 
Here $N^{\text{nos}}$ is the number of non-oscillated  $B$ decays and
$N^{\text{osc}}$ is the number of oscillated $B$ decays.
An event $B \to \mu^+ \nu \adzero X$ with $q(\mu) \times d<0$ 
was tagged as non-oscillated,
and an event with $q(\mu)\times d>0$
was tagged as oscillated.
The flavor tagging variable $d$ is defined in Eq.\ (\ref{ctag1}) 
or (\ref{eq:d_likelihoods}).

All events in the \dzero and \dst samples  were divided into seven groups
according to the measured VPDL ($L$) defined in Eq.\ (\ref{vpdl}). 
The numbers of oscillated $N^{\text{osc}}_i$ and
non-oscillated $N^{\text{nos}}_i$ signal events in each group $i$ were
determined from the number of the \dzero~signal events given by a fit 
to the $K\pi$ invariant mass distribution for both samples. 
The seven VPDL bins (in cm) defined were: \\
$-0.025 < {L} \leq 0.0$, $0.0 < {L} \leq 0.025$, $0.025 < {L} \leq 0.050$, $0.050 < {L} \leq 0.075$, $0.075 < {L} \leq 0.1$, $0.1 < {L} \leq 0.125$ and $0.125 < {L} \leq 0.2$.

\subsection{Mass Fit}
\label{sec:mass-fit}
In this section we describe the mass fitting procedure.
The fitting function was chosen to give the best $\chi^2$ of the fit to
the $K\pi$ mass spectrum of the entire sample of $B \to \mu^+ \adzero X$
events shown in Figs. \ref{fig:fig1} and \ref{fig:fig2}. The signal
peak corresponding to the decay $D^0 \to K^- \pi^+$ can be seen
at 1.857 GeV/$c^2$. The background to the right of the signal region is adequately 
described by an exponential function:

\begin{equation}
f_{1}^{\text {bkg}}(x)   =  a_0 \times e^{-\frac{x}{b_0}},
\label{funcbkg_1}
\end{equation}
where $x$ is the $K\pi$ mass.

The peak in the background to the left of the signal is due to
events in which $D$ mesons decay to  $K\pi X$ where
$X$ is not reconstructed. 
It was modeled with a
bifurcated Gaussian function:

\begin{eqnarray}
f_{2}^{\text{bkg}}(x)  & = & A~ e^{-\frac{(x-\mu_0)^2}{(2 \sigma_R^{2})}}
~~\mbox{for}~~ x-\mu_0 \geq 0 \\
        & = & A~ e^{-\frac{(x-\mu_0)^2}{(2 \sigma_L^{2})}}
~~\mbox{for}~~ x-\mu_0 < 0.  \nonumber
\label{funcbkg_2}
\end{eqnarray}
Here 
$\mu_0$ is the mean
of the Gaussian, and $\sigma_L$ and $\sigma_R$ are the two widths
of the bifurcated Gaussian function.

The signal has been modeled by the sum of two Gaussians:
\begin{eqnarray}
\label{funcsig}
\hspace*{-0.7cm}
f^{\text {sig}} (x) & = & \frac{N^{\text {sig}}}{\sqrt{2 \pi}}  
 \left( \frac{r_1}{\sigma_1} 
e^{-\frac{(x-\mu_1)^2}{2 \sigma_1^2}}
+     \frac{1-r_1}{\sigma_2} 
e^{-\frac{(x-\mu_2)^2}{2 \sigma_2^2}} \right),
\end{eqnarray}
where $N^{\text {sig}}$ is the number of signal events, 
$\mu_1$ and $\mu_2$ are the means of the Gaussians, 
$\sigma_1$ and $\sigma_2$ are the widths of the Gaussians,
and $r_1$ is the fractional contribution of the first Gaussian. 

The complete fitting function, which has twelve free parameters, is:
\begin{equation}
\label{funcfit}
f(x) = f^{\text {sig}}(x) + f_{1}^{\text {bkg}}(x) + f_{2}^{\text{bkg}}(x).
\end{equation}

The low statistics in some VPDL bins, 
which have as few as ten events after flavor tagging, 
do not permit a free fit to this function, 
Consequently some parameters had to be constrained or fixed. In order to do this, 
it was necessary to
show that the constraints on the parameters are valid for all of the VPDL
bins. Unconstrained fits were performed to several high statistic samples, and
the set of all events was used as a reference fit. Events were divided into
VPDL bins and fit to investigate the VPDL dependance of the fit results. 
In addition, three
samples were made to test whether the presence of a flavor
tag changes the mass spectrum: all tagged events over the entire
VPDL range, all events in the short VPDL range [0,0.05] tagged
as opposite-sign events, and all events in VPDL range
[0,0.05] tagged as same-sign events.

\begin{figure}[tbh]
\includegraphics[width=8.8cm]{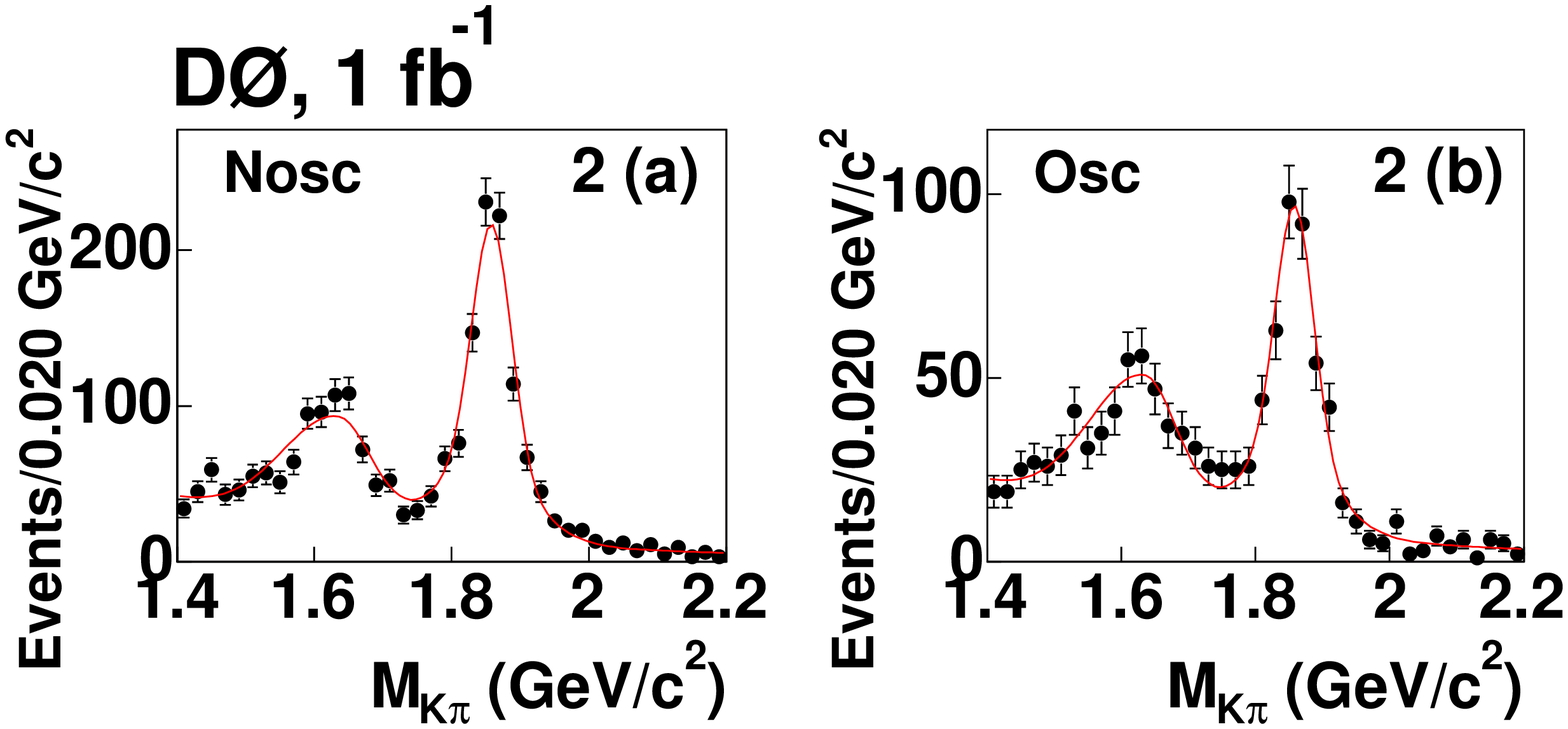} \\
\includegraphics[width=8.8cm]{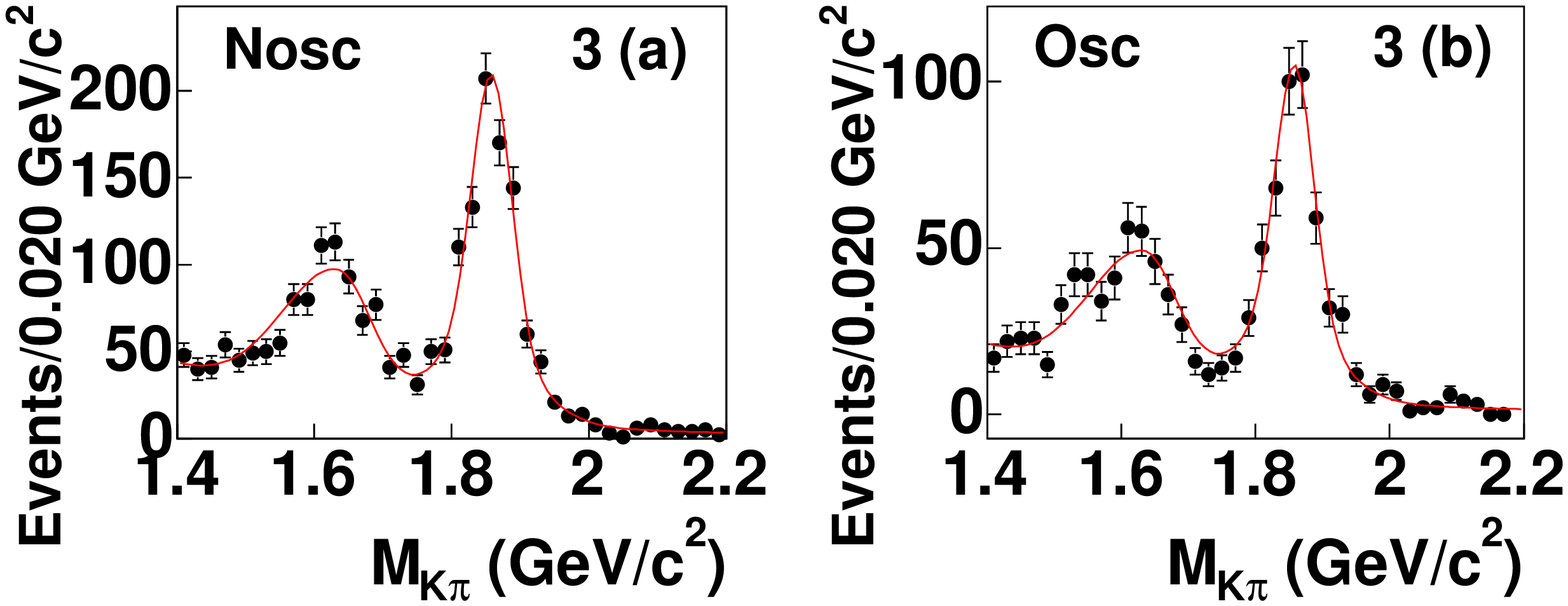} \\
\includegraphics[width=8.8cm]{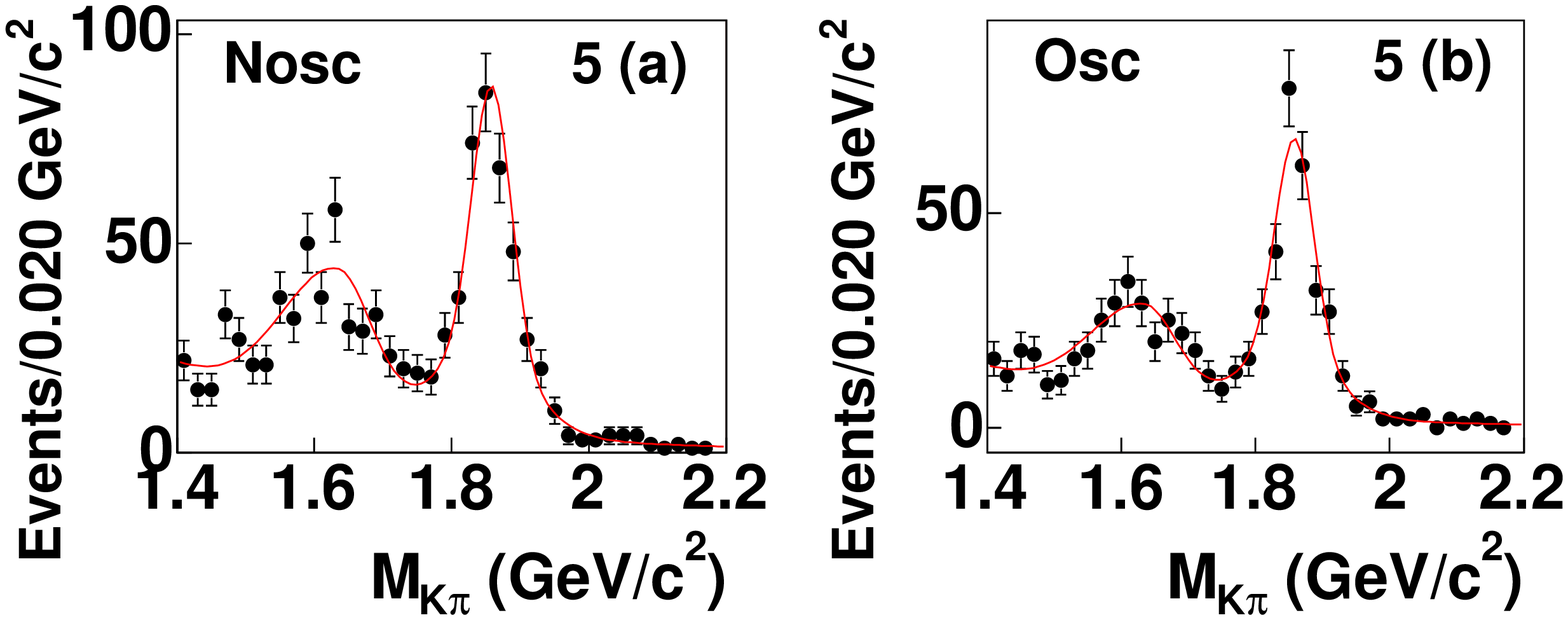} \\
\includegraphics[width=8.8cm]{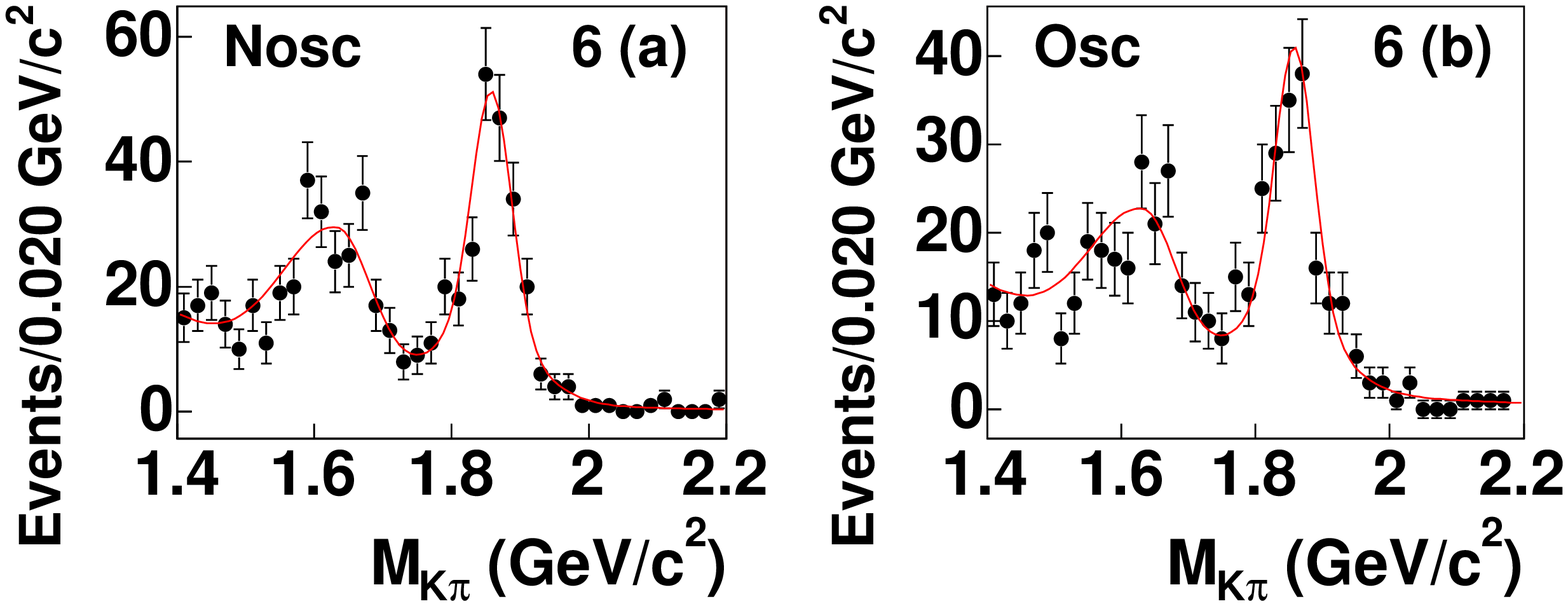} \\
\caption{The fit to  $M(K \pi)$ mass for non-oscillating (left) and 
oscillating (right) $B \to \mu^+ \nu \dstminus$ events tagged by 
the muon tagger with $|d|>0.3$ in VPDL bins 
$0.0 < L \leq 0.025$ cm (2a,2b), 
$0.025 < L \leq 0.050$ cm (3a,3b),
$0.075 < L \leq 0.100$ cm (5a,5b),
and, $0.100 < L \leq 0.125$ cm (6a,6b).
\label{mufit}}
\end{figure}

This study showed that the width, position, and the ratio of the signal 
Gaussians, as well as the position and widths of the bifurcated
Gaussian describing the background can be fixed
to the values obtained from the fit to the 
total \dzero or \dst mass distribution.
This left four free parameters: the
numbers of events in the signal peak, background peak, and
exponential background, and the slope constant of the exponential
background. 
Examples of the fits to the $K \pi$ mass distribution in different
VPDL bins are shown in Fig. \ref{mufit}. 

The number of $D^*$ candidates was estimated using the distribution of
$(M_{K^+ \pi^{-} \pi^{-}} - M_{K^{+} \pi^{-}})$, 
shown in Fig. \ref{fig:fig3}.
In this case, the signal was modeled with two Gaussians as described by
Eq.(\ref{funcsig}), and the background
by the product of a linear and exponential function
\begin{eqnarray}
f^{\text{bkg}}(x) & = & a \left[ 1 + c (x - x_0) \right] e^{\frac{x-x_0}{b_0}}.
\label{funcdst}
\end{eqnarray}
where $x$ is the mass difference $(M_{K^+ \pi^{-} \pi^{-}} - M_{K^{+} \pi^{-}})$ in this equation.

\subsection{Expected flavor asymmetry}
\label{asymfit}

For a given type of $B_q$ meson ($q= u, d, s$), the distribution of the
visible proper decay length $L$ is given by 

\begin{eqnarray}
n^{\text {nos}}_u(L,K) & = & \frac{1}{2} \cdot \frac{K}{c\tau(B^+)} e^{ \left(-\frac{KL}{c\tau(B^+)} \right)}
           (1+\dilu), \\
n^{\text {osc}}_u(L,K) & = & \frac{1}{2} \cdot \frac{K}{c\tau(B^+)} e^{ \left(-\frac{KL}{c\tau(B^+)} \right)}
           (1-\dilu), \\
n^{\text {nos}}_d(L,K) & = & \frac{1}{2} \cdot \frac{K}{c\tau(B^0)} e^{ \left(-\frac{KL}{c\tau(B^0)} \right)} \cdot \nonumber \\
                       &  & \left [ 1+\dild \cos(\dmd \frac{K L}{c}) \right ], \\
n^{\text {osc}}_d(L,K) & = & \frac{1}{2} \cdot \frac{K}{c\tau(B^0)} e^{ \left(-\frac{KL}{c\tau(B^0)} \right)} \cdot \nonumber \\
                       & &  \left [1-\dild \cos(\dmd \frac{K L}{c}) \right], \\
n^{\text{nos}}_s(L,K) & = & \frac{1}{2} \cdot \frac{K}{c\tau(B^0_s)} e^{ \left(-\frac{KL}{c\tau(B^0_s)} \right)}, \\
n^{\text{osc}}_s(L,K) & = & \frac{1}{2} \cdot \frac{K}{c\tau(B^0_s)} e^{\left(-\frac{KL}{c\tau(B^0_s)} \right)}.
\end{eqnarray}

Here $\tau$ is the lifetime of the $B$ meson, \dmd~ is the mixing frequency
of $B^0$ mesons, the factor $K = P_T^{\mu \dzero}/P_T^B$
reflects the difference between the measured ($P_T^{\mu \dzero}$)
and true ($P_T^B$) momenta of the $B$ meson.
The $B^+$ meson does not oscillate, and it is assumed in these studies that 
the $B^0_s$ meson oscillates with infinite frequency. 
The flavor tagging dilution is given by ${\cal D}$. In general, it can be different for $B^0$ and $B^+$. In our study we verified the 
assumption that $\dild = \dilu$ for our opposite-side flavor tagging.

The transition from the true to the experimentally measured visible
proper decay length $L^M$ is achieved by integration 
over the $K$-factor distribution and convolution with the resolution function: 

\begin{eqnarray}
N^{\text {nos/osc}}_{q,~j}(L^M) & = & \int dL~ R_j(L-L^M)~
\varepsilon_j(L)~ \theta(L) \nonumber \\
 & \times & \int dK~ D_j(K) ~n^{\text {nos/osc}}_{q,~j}(L,K) 
\end{eqnarray}

Here $R_j(L-L^M)$ is the detector resolution in the VPDL, and 
$\varepsilon_j(L)$ is the reconstruction efficiency for a given channel 
$j$ of $B_q$ meson decay. The step function $\theta(L)$ forces $L$ to be positive
in the integration. $L^M$ can be negative due to resolution effects.
The function $D_j(K)$ is a normalized distribution of the $K$-factor
in a given channel $j$, obtained from simulated events.

In addition to the main decay channel $B \to \mu^+ \nu \adzero X$, the process
$c \bar c \to \mu^+ \nu \adzero X$ contributes to the selected final state.
A dedicated analysis was developed to study this process, both in data and in simulation.
It shows that the pseudo decay-length, constructed from
the crossing of the $\mu$ and \adzero~ trajectories, is distributed around zero
with $\sigma \approx 150~\mu$m. The distribution $N^{c\bar{c}} (L^M)$ of the VPDL
for this process was taken from simulation. It was assumed
that the production ratio $(c \to D^*)/(c \to D^0)$ is the same as in semileptonic
$B$ decays and that the flavor tagging for the $c \bar c$ events gives the
same rate of oscillated and non-oscillated events. The fraction $f_{c\bar{c}}$
of $c \bar c$ events was obtained from the fit.

Taking into account all of the above mentioned contributions, 
the expected number of (non-) oscillated events in the $i$-th bin of VPDL is

\begin{eqnarray}
N^{e,\text {nos/osc}}_i  & = & \int_i dL^M~  (1-f_{c \bar c}) \nonumber \\
& \times & \left( \sum_{q=u,d,s} \sum_j \text{Br}_j \cdot N^{\text {nos/osc}}_{q,~j}(L^M) \right) \nonumber \\
             & + & \int_i dL^M~ f_{c \bar{c}} N_{c\bar{c}}(L^M).
\end{eqnarray}
Here the integration $\int_i dL^M$ is taken over a given interval $i$,
the sum $\sum_j$ is taken over all decay channels $B_{q} \to \mu^+ \nu \adzero X$
contributing to the selected sample,
and $\text{Br}_j$ is the branching fraction of channel $j$.

Finally, the expected value of asymmetry, $A_i^e(\Delta m, f_{c\bar{c}}, \dild, \dilu)$, for the interval $i$ of the measured VPDL is given by

\begin{equation}
\label{asymexp}
A_i^e(\Delta m, f_{c\bar{c}}, \dild, \dilu) = \frac{N^{e,\text {nos}}_i - N^{e,\text{osc}}_i}{N^{e,\text{nos}}_i + N^{e,\text{osc}}_i}.
\end{equation}

The expected asymmetry can be computed both for the \dst and the \dzero samples.
The only difference between them is due to the different relative contributions of various decay channels of $B$ mesons.

For the computation of $A_i^e$, the $B$ meson lifetimes and
the branching fractions $\text{Br}_j$ were taken from the Particle Data Group (PDG) \cite{pdg}. 
They are discussed in the following section. The functions $D_j(K)$, $R_j(L)$,
and $\varepsilon_j(L)$ were obtained from MC simulation. Variations of these inputs 
within their uncertainties are included in the systematic uncertainties.

\subsection{Sample Composition}
\label{sec:istbd}
There is a cross-contamination between the $b \rightarrow B^0\rightarrow \mu^+ \nu \adzero X$, $b \rightarrow B^0_s \to \mu^+ \nu \adzero X$;  and $b \rightarrow B^+\rightarrow \mu^+ \nu \adzero X$  samples.
To determine the composition of the selected samples, we studied all possible decay chains for $B^0$, $B^0_s$, and $B^+$ with their corresponding branching fractions, from which we estimated the sample composition in the $D^*$ and $D^0$ samples.

The following decay
channels of $B$ mesons were considered for the $D^*$ sample:
\begin{eqnarray}
B^0 & \to & \mu^+ \nu \dstminus, \nonumber \\
B^0 & \to & \mu^+ \nu \dststminus \to \mu^+ \nu \dstminus X, \nonumber \\
B^+ & \to & \mu^+ \nu \adststzero \to \mu^+ \nu \dstminus X, \nonumber \\
B^0_s & \to & \mu^+ \nu \dstminus X; \nonumber 
\end{eqnarray}
and for the \dzero~ sample:
\begin{eqnarray}
B^+ & \to & \mu^+ \nu \adzero, \nonumber \\
B^+ & \to & \mu^+ \nu \adstzero, \nonumber \\
B^+ & \to & \mu^+ \nu \adststzero  \to  \mu^+ \nu \adzero X, \nonumber \\
B^+ & \to & \mu^+ \nu \adststzero \to \mu^+ \nu \adstzero X, \nonumber \\
B^0 & \to & \mu^+ \nu \dststminus \to \mu^+ \nu \adzero X, \nonumber \\
B^0 & \to & \mu^+ \nu \dststminus \to \mu^+ \nu \adstzero X, \nonumber \\
B^0_s & \to & \mu^+ \nu \adzero X, \nonumber \\
B^0_s & \to & \mu^+ \nu \adstzero X. \nonumber 
\end{eqnarray}
Here, and in the following, the symbol ``$D^{**}$'' denotes both narrow
and wide $D^{**}$ resonances, as well as non-resonant
$\dmes \pi$ and $\dst \pi$ production.

The most recent PDG values~\cite{pdg} were used to
determine the branching fractions of decays
contributing to the $D^0$ and $D^*$ samples:
\begin{eqnarray}
{\rm Br}(B^+ \to \mu^+ \nu \adzero) & = & (2.15 \pm 0.22)\times 10^{-2}, \nonumber \\
{\rm Br}(B^0 \to \mu^+ \nu \dmes^-) & = & (2.14 \pm 0.20)\times 10^{-2}, \nonumber \\
{\rm Br}(B^+ \to \mu^+ \nu \adstzero) & = & (6.5 \pm 0.5)\times 10^{-2}, \nonumber \\
{\rm Br}(B^0 \to \mu^+ \nu \dstminus) & = & (5.44 \pm 0.23)\times 10^{-2}. \nonumber
\end{eqnarray}

Br$(B^+ \to \mu^+ \nu \adststzero)$ 
was estimated using the following inputs:
\begin{eqnarray}
{\rm Br}(B\to \mu^+ \nu X) & = & (10.73 \pm 0.28)\times 10^{-2}~\mbox{\cite{pdg}}, \nonumber \\
{\rm Br}(B^0 \to \mu^+ \nu X) & = & \tau(B^0)/\tau(B^+) \nonumber \\
&  \cdot & {\rm Br} (B^+ \to \mu^+ \nu X), \nonumber \\
{\rm Br}(B^+ \to \mu^+ \nu \adststzero) & = & {\rm Br}(B^+ \to \mu^+ \nu X)  \nonumber \\
& - & {\rm Br}(B^+ \to \mu^+ \nu \adzero)  \nonumber \\
& - & {\rm Br}(B^+ \to \mu^+ \nu \adstzero), 
\end{eqnarray}
where $\tau(B^0)$ is the $B^0$ lifetime, and $\tau(B^+)$ is the $B^+$ lifetime.
The following value was obtained:
\begin{equation}
\mbox{Br}(B^+ \to \mu^+ \nu \adststzero) = (2.70 \pm 0.47)\times 10^{-2}. 
\end{equation}

\noindent
${\rm Br}(B^0 \to \mu^+ \nu \dststminus)$ is obtained as follows:
\begin{eqnarray}
{\rm Br}(B^0 \to \mu^+ \nu \dststminus) & = & {\rm Br}(B^0 \to \mu^+ \nu X)  \nonumber \\
& - & {\rm Br}(B^0 \to \mu^+ \nu \dminus)  \nonumber \\
& - & {\rm Br}(B^+ \to \mu^+ \nu \dstminus), \nonumber \\
{\rm Br}(B^0 \to \mu^+ \nu \dststminus) & = & \frac{\tau(B^0)}{\tau(B^+)} \cdot
      {\rm Br} (B^+ \to \mu^+ \nu \adststzero) \nonumber
\end{eqnarray}

\noindent
Br$(B^+ \to \mu^+ \nu \adststzero \to \mu^+ \nu \dstminus X)$
was estimated from the following inputs:
\begin{eqnarray}
{\rm Br}(\bar{b} \to l^+ \nu \dstminus \pi^+ X) &=&  (4.73 \pm 0.8 \pm 0.6) \times 10^{-3}  ~\mbox{\cite{ALEPH}},  \nonumber \\
{\rm Br}(\bar{b} \to l^+ \nu \dstminus \pi^+ X) &=&  (4.80 \pm 0.9 \pm 0.5) \times 10^{-3} ~\mbox{\cite{DELPHI}}, \nonumber \\
{\rm Br}(\bar{b} \to l^+ \nu \dstminus \pi^- X) &=&  (0.6 \pm 0.7 \pm 0.2) \times 10^{-3} ~\mbox{\cite{DELPHI}} \nonumber 
\end{eqnarray}
and assuming \mbox{Br}$(b \to B^+) = (0.397 \pm 0.010)$~\cite{pdg}. The usual practice in estimating this decay rate is to neglect 
the contributions of the decays $D^{**}\to D^*\pi\pi$. However, the above data allows us to take these decays into account.

Neglecting the decays $D^{**}\to D^*\pi\pi\pi$, 
the available measurements can be expressed as:
\begin{eqnarray}
\mbox{Br}(\bar{B} \to l^+ \nu \dstminus \pi^+ X) & = &
\mbox{Br}(B^+ \to l^+ \nu \dstminus \pi^+ X^0), \nonumber \\
& + & \mbox{Br}(B^0 \to l^+ \nu \dstminus \pi^+ \pi^-), \nonumber \\
\mbox{Br}(\bar{B} \to l^+ \nu \dstminus \pi^- X) & = & 
\mbox{Br}(B^0 \to l^+ \nu \dstminus \pi^+ \pi^-). \nonumber
\end{eqnarray}
From these relations and using the above measurements, we obtain
\begin{eqnarray}
\mbox{Br}(B^+ & \to & \mu^+ \nu \adststzero \to l^+ \nu \dstminus X) = \nonumber \\
& & (1.06 \pm 0.24)\times 10^{-2}.
\end{eqnarray}

All other factors for the $\mbox{Br}(B \to \mu^+ \nu \adstst \to \mu^+ \nu \adst X)$
were obtained assuming the following relations,
\begin{eqnarray}
\frac{{\rm Br}(B^0 \to \mu^+ \nu \dststminus \to \mu^+ \nu \dst \pi )}
     {{\rm Br}(B^+ \to \mu^+ \nu \adststzero \to \mu^+ \nu \dst \pi )}
& = & \tau(B^0)/\tau(B^+), \nonumber \\
\frac{{\rm Br}(B \to \mu^+ \nu \adstst \to \mu^+ \nu \adst \pi^+ )}
     {{\rm Br}(B \to \mu^+ \nu \adstst \to \mu^+ \nu \adst \pi^0 )}
& = & 2. \nonumber
\end{eqnarray}

\noindent
${\rm Br}(B \to \mu^+ \nu \adstst \to \mu^+ \nu \ad X)$ was estimated
from the following inputs:
\[
\frac{{\rm Br}(B \to \mu^+ \nu \adstst \to \mu^+ \nu \ad \pi^+)}
     {{\rm Br}(B \to \mu^+ \nu \adstst \to \mu^+ \nu \ad \pi^0)} = 2,
\]
\begin{eqnarray}
{\rm Br}(B \to \mu^+ \nu \adstst ) & = & 
{\rm Br}(B \to \mu^+ \nu \adstst \to \mu^+ \nu \ad X) \nonumber \\
& + & {\rm Br}(B \to \mu^+ \nu \adstst \to \mu^+ \nu \adst X). \nonumber
\end{eqnarray}

\noindent
To estimate branching fractions for $B_s^0$ decays, 
${\rm Br}(B^0_s \to \mu^+ \nu D_s^- X) = (7.9 \pm 2.4)\times 10^{-2}$ was taken 
from Ref. \cite{pdg} and  the following assumptions were used:
\begin{eqnarray}
& & \frac{{\rm Br}(B^0_s \to \mu^+ \nu X) }
     {{\rm Br}(B^0 \to \mu^+ \nu X)}  =  \tau(B_{s}^{0})/\tau(B^0), \nonumber \\
& & \frac{{\rm Br}(B^0_s \to \mu^+ \nu D_s^{**-} \to \mu^+ \nu \dstminus X)}
     {{\rm Br}(B^0_s \to \mu^+ \nu D_s^{**-} \to \mu^+ \nu \adstzero X)} 
  =  1, \nonumber
\end{eqnarray}
where $\tau(B_{s}^{0})$ is the $B^0_s$ meson lifetime. In addition, it was assumed that
\begin{equation}
\frac{\mbox{{\rm Br}}(B^0_s \to \mu^+ \nu D_s^{**-} \to \mu^+ \nu \dst X)}
       {\mbox{{\rm Br}}(B^0_s \to \mu^+ \nu D_s^{**})} = 0.35.
\label{rs}
\end{equation}
There is no experimental measurement of this ratio yet and to estimate the
the corresponding systematic uncertainty, this ratio was varied between
0 and 1.

In addition to these branching fractions, 
various decay chains are affected differently by the 
$B$ meson selection cuts, and 
the corresponding reconstruction efficiencies were determined from
simulation to correct for this effect.
Taking into account these efficiencies, the composition of the
\dst~ sample was estimated to be $(0.89 \pm 0.03)$ $B^0$,
$(0.10 \pm 0.03)$ $B^+$, and $(0.01 \pm 0.01)$ $B^0_s$. The \dzero sample contains
$(0.83 \pm 0.03)$ $B^+$, $(0.16 \pm 0.04)$ $B^0$, and $(0.01 \pm 0.01)$ $B^0_s$.

Since the \dst sample was selected by the cut on the mass difference 
$\Delta M = M(\dzero \pi) - M(\dzero)$, there is a small 
additional contribution 
of $B \to \mu^+ \nu \adzero$ events to the \dst sample, when \dzero
is randomly combined with a pion from the combinatorial background.
The fraction of this contribution was estimated using $\mu^+ \adzero \pi^+$ events.
These events were selected applying all the criteria for the \dst sample,
described in Sec. \ref{sec:measmeth}, except that the
wrong charge correlation of muon and pion was required, i.e., the
muon and the pion were required to be having the same charge.
The number of \dzero events was determined using the same fitting procedure
as for the \dst sample, and the additional fraction of  $B \to \mu^+ \nu \adzero$ 
events in the \dst sample was estimated to be $(4.00 \pm 0.85)\times 10^{-2}$.
This fraction was included in the fitting procedure and 
the uncertainty in this value was taken into account in the overall systematics.

\section{Results}
\label{sec:obs}
\begin{figure*}
\includegraphics[width=11.0cm]{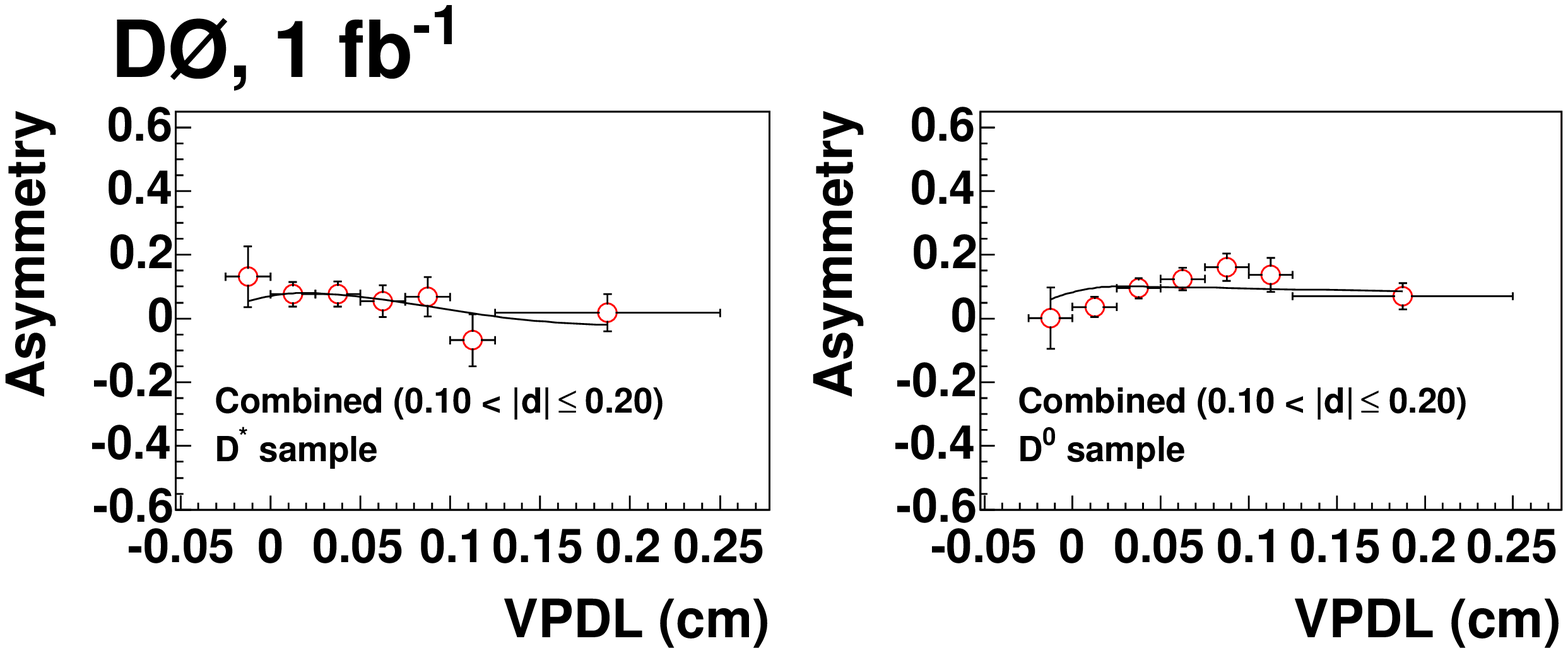} 
\includegraphics[width=11.0cm]{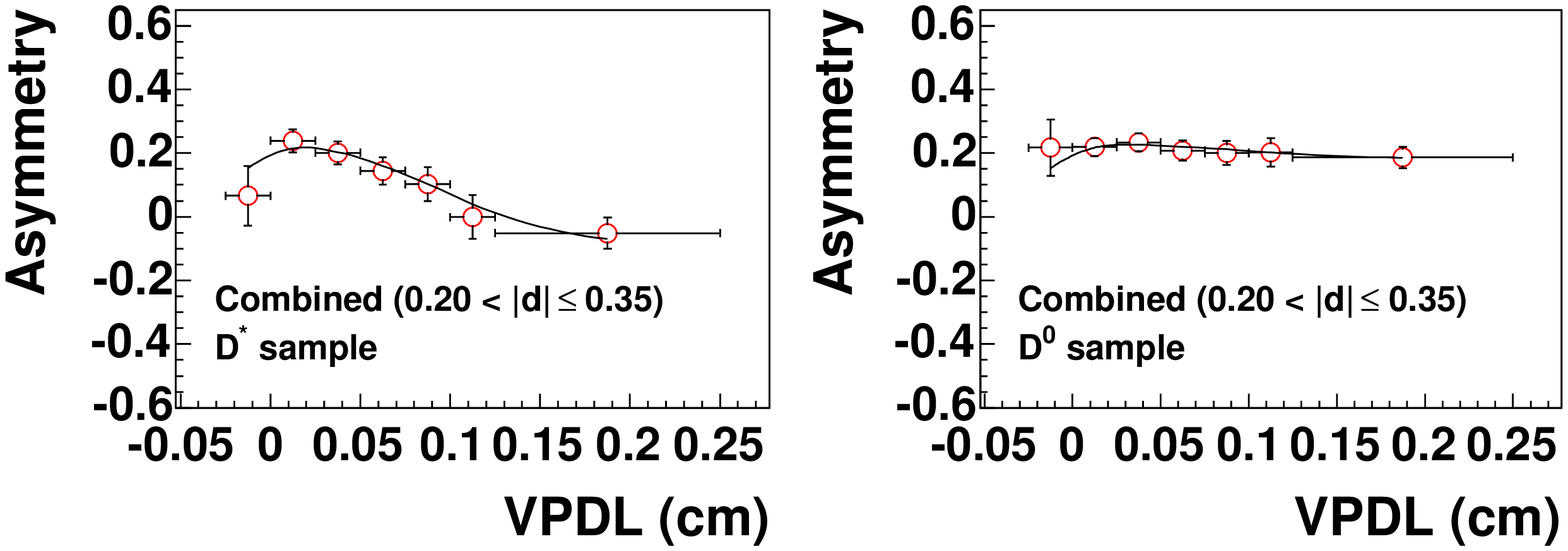}
\includegraphics[width=11.0cm]{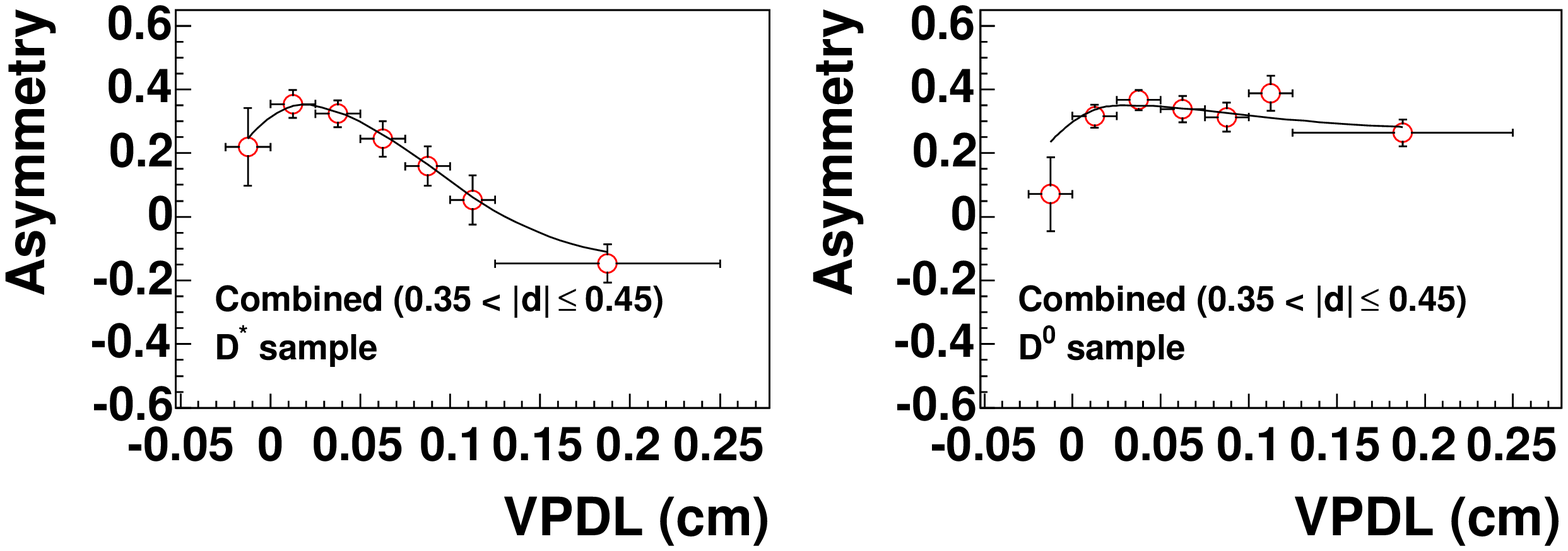} 
\includegraphics[width=11.0cm]{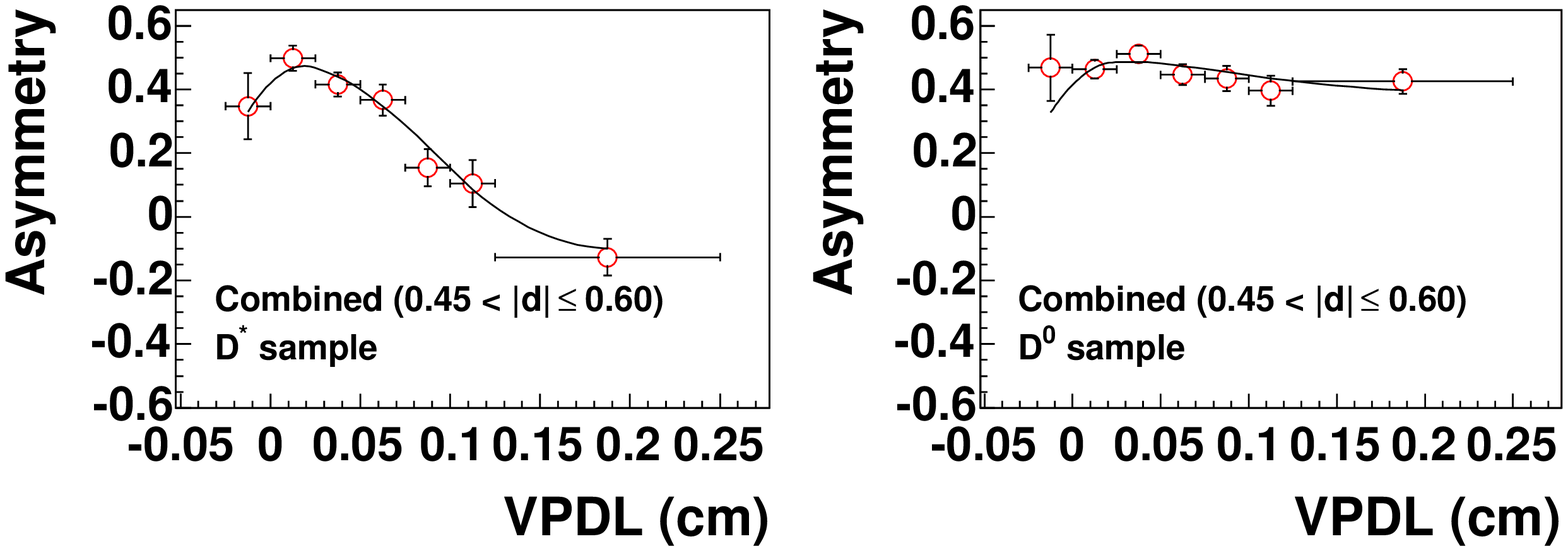}
\includegraphics[width=11.0cm]{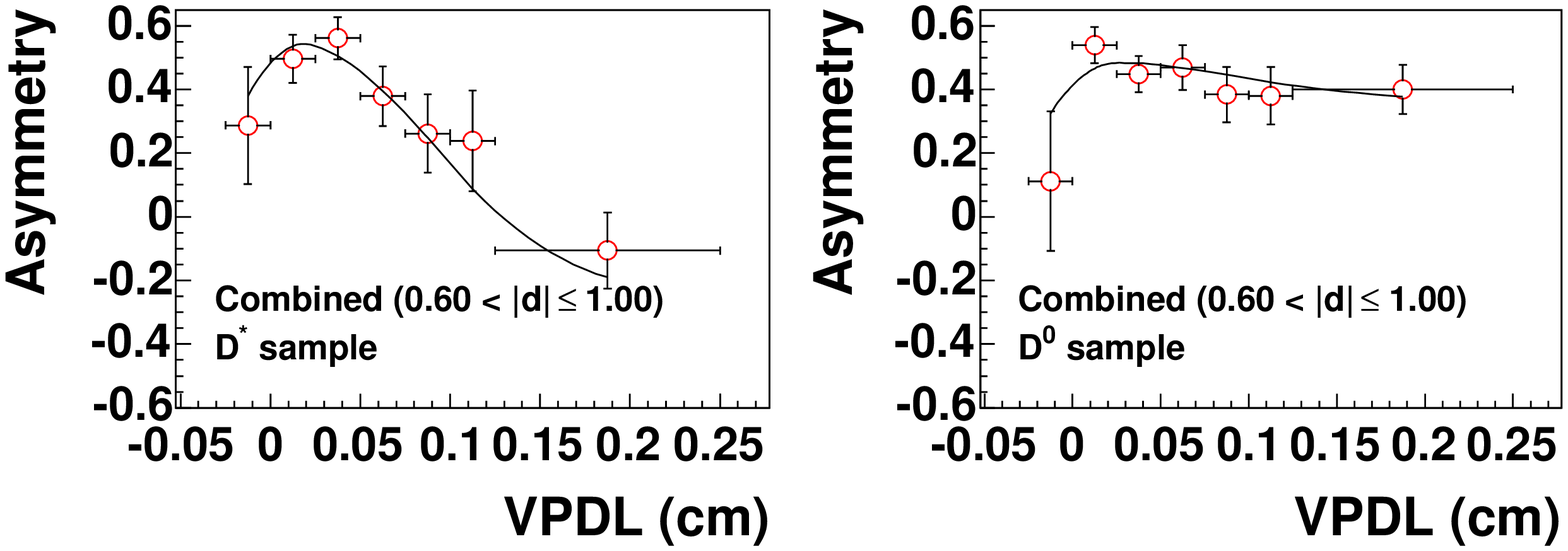}
\caption{Asymmetries obtained in the \dst and \dzero sample with the 
combined tagger in $|d|$ bins.
Circles are data, and the result of the fit 
is superimposed.
\label{asymfits1}}
\end{figure*}

For each sample of tagged events, the observed and expected asymmetries
were determined using Eqs. (\ref{asymobs}) and (\ref{asymexp}) in all VPDL bins, 
and the values of $\dmd$, $f_{c\bar{c}}$, $\dilu$, and $\dild$
were obtained from a simultaneous $\chi^2$ fit: 
\begin{align}
\label{chi2_1}
\lefteqn{\chi^2(\dmd, f_{c\bar{c}},\dild,\dilu) = } \nonumber \\
 & \chi^2_{\dst}(\dmd, f_{c\bar{c}},\dild,\dilu)  + 
   \chi^2_{\dzero}(\dmd, f_{c\bar{c}},\dild,\dilu)  ; \\
\lefteqn{\chi^2_{\dst}(\dmd, f_{c\bar{c}},\dild,\dilu) = } \nonumber \\
 & \sum_i \frac{[A_{i,\dst} - A_{i,\dst}^e(\dmd, f_{c\bar{c}},\dild, \dilu)]^2}{\sigma^2(A_{i,\dst})};   \\
\lefteqn{\chi^2_{\dzero}(\dmd, f_{c\bar{c}},\dild, \dilu)  = } \nonumber \\
 & \sum_i 
\frac{[A_{i,\dzero} - A_{i,\dzero}^e(\dmd, f_{c\bar{c}},\dild, \dilu)]^2}{\sigma^2(A_{i,\dzero})}.
\end{align}
Here $\sum_i$ is the sum over all VPDL bins.
Examples of the $\chi^{2}$ fit to the flavor asymmetry minimization given
in Eq. (\ref{chi2_1}), is shown in Fig. \ref{asymfits1}.

The performance of the flavor tagging method was studied 
separately for the muon, electron, and secondary vertex taggers
using events with $|d| > 0.3$. Results 
are given in Tables \ref{meritd}--\ref{Dmfits}. All uncertainties in these
tables are statistical only and do not
include systematic uncertainties. The performances of the combined
tagger defined in Sec.\ref{sec:combined} for events with $|d| > 0.3$
and the alternative multidimensional tagger defined in Sec.\ref{sec:mdimtag} 
for events with $|d| > 0.37$ are also shown. The cut on $|d|$ is somewhat
different for the multidimensional tagger as the calibration is different
and we compare the dilutions for the same tag efficiency for the two taggers.
The tagging efficiencies shown in Tables \ref{meritd} and \ref{meritu}
were computed using events with VPDL=[0.025,0.250]. This selection
reduces the contribution from $c \bar c \to \mu^+ \nu D^0 X$ events, 
since they have a VPDL distribution with zero mean and $\sigma \approx 150~ \mu$m
as described in Sec.\ref{sec:massfitproc}.

\begin{table*}
\caption{
\label{meritd}
Tagging performance for the \dst sample for different taggers and subsamples. Uncertainties
are statistical only.
}
\begin{ruledtabular}
\begin{tabular}{|l|c|c|c|}
Tagger   & $\varepsilon (\%)$ & \dild & $\varepsilon \dild^2 (\%)$ \\\hline
Muon (${|d|} > 0.3$)        & $\,\,\, 6.61 \pm 0.12$  & $0.473 \pm 0.027$ & $1.48 \pm 0.17$  \\
Electron (${|d|} > 0.3$)    & $\,\,\, 1.83 \pm 0.07$  & $0.341 \pm 0.058$ & $0.21 \pm 0.07$  \\
SVCharge (${|d|} > 0.3$)    & $\,\,\, 2.77 \pm 0.08$  & $0.424 \pm 0.048$ & $0.50 \pm 0.11$  \\
Combined (${|d|} > 0.3$)    & $11.14 \pm 0.15$ & $0.443 \pm 0.022$  & $2.19 \pm 0.22$  \\
Multidim. (${|d|} > 0.37$)   & $10.98 \pm 0.15$  & $0.395 \pm 0.022$ & $1.71 \pm 0.19$ \\
Combined (0.10$<{|d|} \leq $0.20) & $\,\,\, 4.63\pm 0.10$  & $0.084 \pm 0.031$ & $0.03 \pm 0.02$ \\
Combined (0.20$<{|d|} \leq $0.35) & $\,\,\, 5.94\pm 0.12$  & $0.236 \pm 0.027$ & $0.33 \pm 0.08$  \\
Combined (0.35$<{|d|} \leq $0.45) & $\,\,\, 3.89\pm 0.09$  & $0.385 \pm 0.034$ & $0.58 \pm 0.10$  \\
Combined (0.45$<{|d|} \leq $0.60) & $\,\,\, 4.36\pm 0.10$  & $0.512 \pm 0.032$ & $1.14 \pm 0.14$  \\
Combined (0.60$<{|d|} \leq $1.00) & $\,\,\, 1.13\pm 0.05$  & $0.597 \pm 0.058$ & $0.40 \pm 0.08$  \\
\end{tabular}
\end{ruledtabular}
\end{table*}

\begin{table*}
\caption{
\label{meritu}
Tagging performance for the \dzero sample for different taggers and subsamples. 
For comparison, the dilution ${\cal D}'_d$ measured in the \dst sample 
with addition
of wrong sign $\mu^+ \nu \adzero \pi^+$ events is also shown. Uncertainties are statistical
only.
}
\begin{ruledtabular}
\begin{tabular}{|l|c|c|c|c|}
Tagger   & $\varepsilon (\%)$ & \dilu & $\varepsilon \dilu^2 (\%)$  & ${\cal D}'_d$ \\\hline
Muon (${|d|} > 0.3$)        & $\,\,\, 7.10 \pm 0.09$  & $0.444 \pm 0.015$ &  $1.400 \pm 0.096$ & $0.463 \pm 0.028$ \\
Electron (${|d|} > 0.3$)    & $\,\,\,1.88 \pm 0.05$  & $0.445 \pm 0.032$ &  $0.372 \pm 0.054$ & $0.324 \pm 0.060$ \\
SVCharge (${|d|} > 0.3$)    & $\,\,\,2.81 \pm 0.06$  & $0.338 \pm 0.026$ &  $0.320 \pm 0.050$ & $0.421 \pm 0.049$ \\
Combined (${|d|} > 0.3$)    & $11.74 \pm 0.11$  & $0.419 \pm 0.012$ &  $2.058 \pm 0.121$ & $0.434 \pm 0.023$ \\
Multidim. (${|d|} > 0.37$)   & $11.67 \pm 0.11$  & $0.363 \pm 0.012$ &  $1.540 \pm 0.106$ & $0.384 \pm 0.023$ \\
Combined (0.10$<{|d|} \leq $0.20) & $\,\,\,4.59 \pm 0.08$  & $0.104 \pm 0.017$ &  $0.050 \pm 0.016$ & $0.079 \pm 0.029$ \\
Combined (0.20$<{|d|} \leq $0.35) & $\,\,\,6.10 \pm 0.09$  & $0.234 \pm 0.014$ &  $0.335 \pm 0.042$ & $0.212 \pm 0.024$ \\
Combined (0.35$<{|d|} \leq $0.45) & $\,\,\,3.98 \pm 0.07$  & $0.361 \pm 0.018$ &  $0.519 \pm 0.052$ & $0.364 \pm 0.032$ \\
Combined (0.45$<{|d|} \leq $0.60) & $\,\,\,4.77 \pm 0.07$  & $0.504 \pm 0.016$ &  $1.211 \pm 0.077$ & $0.489 \pm 0.030$ \\
Combined (0.60$<{|d|} \leq $1.00) & $\,\,\,1.17 \pm 0.04$  & $0.498 \pm 0.031$ &  $0.290 \pm 0.038$ & $0.572 \pm 0.056$ \\
\end{tabular}
\end{ruledtabular}
\end{table*}

\begin{table}[htbp]
\caption{
\label{Dmfits}
Measured value of $\dmd$ and $f_{c \bar c}$ for different taggers 
and subsamples.
}
\begin{ruledtabular}
\begin{tabular}{|l|c|c|}
Tagger & $\dmd$ (ps$^{-1}$) & $f_{c \bar c}$ \\\hline
Muon                       & $0.502 \pm 0.028$ & $0.013 \pm 0.010$ \\
Electron                   & $0.481 \pm 0.067$ & $0.058 \pm 0.045$ \\
SVCharge                  & $0.553 \pm 0.053$ & $0.096 \pm 0.050$ \\
Multidim.                   & $0.502 \pm 0.026$ & $0.031 \pm 0.014$ \\
Combined (${|d|}>0.3$)      & $0.513 \pm 0.023$ & $0.033 \pm 0.013$ \\
Combined ($0.10 < {|d|} \leq 0.20$) & $0.506 \pm 0.209$ & $0.495 \pm 0.505$  \\
Combined ($0.20 < {|d|} \leq 0.35$) & $0.523 \pm 0.064$ & $0.021 \pm 0.025$  \\
Combined ($0.35 < {|d|} \leq 0.45$) & $0.531 \pm 0.042$ & $0.063 \pm 0.038$ \\
Combined ($0.45 < {|d|} \leq 0.60$) & $0.510 \pm 0.032$ & $0.010 \pm 0.010$ \\
Combined ($0.60 < {|d|} \leq 1.00$) & $0.456 \pm 0.049$ & $0.032 \pm 0.026$  \\
\end{tabular}
\end{ruledtabular}
\end{table}

Individual taggers give compatible values of \dmd~
and $f_{c \bar c}$, as can be seen in Table~\ref{Dmfits}.
 For the combined tagger with $|d|>0.3$, the following results were 
obtained:
\begin{eqnarray}
\label{tagpower1}
\varepsilon \dild^2 & = & (2.19 \pm 0.22) \%, \nonumber \\
\dmd & = & 0.513 \pm 0.023  ~{\rm ps}^{-1}, \nonumber \\
f_{c \bar c} & = & (3.3 \pm 1.3) \%.
\end{eqnarray}
The multidimensional tagger which used simulation for the description
of p.d.f.'s as described in Sec.\ref{sec:mdimtag} gives
consistent results both for the \dmd~ and the fraction $f_{c \bar c}$,
which is used as a cross-check of the main tagging algorithm.

One of the goals of this analysis was to verify the assumption
of independence of the opposite-side flavor tagging on the type of the
reconstructed $B$ meson. It can be seen from Tables \ref{meritd} and \ref{meritu}
that the measured flavor tagging performance for $B^0$ events is slightly
better than for $B^+$ events, both for individual and combined taggers.
This difference can be explained by the better selection of $\mu^+ \nu \dstminus$
events due to an additional requirement of the charge correlation between the 
muon and pion from $\dstminus \to \dzero \pi^-$ decay. The \dzero~
sample can contain events with a wrongly selected muon. Since the
charge of the muon determines the flavor asymmetry, such 
a background can reduce
the measured $B^+$ dilution. The charge correlation between
the muon and the pion suppresses this background and results in a
better measurement of the tagging performance.

To verify this hypothesis, a special sample of events satisfying all conditions
for the \dst sample, except the requirement of the charge correlation between
the muon and the pion, was selected. The dilution ${\cal D}'_d$ for this
sample is shown in Table~\ref{meritu}. It can be seen that ${\cal D}'_d$
is systematically lower than \dild~ for all samples and all taggers.
${\cal D}'_d$ is the right quantity to be compared with \dilu~ and 
Table~\ref{meritu} shows that they are statistically compatible.
This result therefore confirms the expectation of the same performance 
of the opposite-side flavor tagging for $B^+$ and $B^0$ events. 
It also shows that contribution of background 
in the \dzero~ sample reduces the measured dilution for $B^+$ events. 
Thus, the dilution measured in the \dst~ sample can be used for the 
$B^0_s$ mixing measurement, where a similar charge correlation between
the muon and $D_s$ is required.

By construction, the dilution for each event
should strongly depend on the magnitude of the tagging variable $d$.
This property becomes important in the $B^0_s$ mixing measurement, 
since in this case the dilution 
of each event can be estimated using the value of $d$ and 
can be included in a likelihood function, improving 
the sensitivity of the measurement. 
To test the dependence of the dilution on $d$,
all tagged events were divided into subsamples with 
$0.1 < |d| \leq 0.2$, $0.2 < |d| \leq 0.35$, $0.35 < |d| \leq 0.45$, 
$0.45 < |d| \leq 0.6$, and $ |d| > 0.6$. The overall tagging efficiency 
for this sample is $(19.95 \pm 0.21) \%$. The dilutions obtained 
are shown in Table \ref{meritd}. Their strong dependence on the value
of the tagging variable is clearly seen. This allows us to perform 
a dilution calibration and obtain the measured dilution $\dild$ as a function
of the predicted value $|d|$. This is used to provide an event-by-event
dilution for the $B_{s}$ mixing analysis and the calibration derived
in this analysis is used for the two sided C.L. on $B_{s}$ mixing, obtained
by D\O\ ~\cite{bsd0}. The overall tagging power,
computed as the sum of the tagging powers in all subsamples, is:
\begin{equation}
\label{tagpower}
\varepsilon \dild^2 = (2.48 \pm 0.21) \%.
\end{equation}

The measured oscillation parameters \dmd~ for all considered taggers and 
subsamples are given in Table~\ref{Dmfits}.
They are compatible with the world average value 
$\dmd = 0.502 \pm 0.007$ ps$^{-1}$~\cite{pdg} in each instance.

The final mixing parameter \dmd~ was obtained from the simultaneous
fit of the flavor asymmetry in the various tagging variable subsamples 
defined above.
The fraction $f_{c \bar c}$ was constrained to be the same for all subsamples.
The result is
\begin{eqnarray}
\label{result}
\dmd & = & 0.506 \pm 0.020~\mbox{ps}^{-1} \\
f_{c \bar c} & = & (2.2 \pm 0.9) \%. \nonumber
\end{eqnarray}

The statistical precision of $\dmd$ from the simultaneous fit is about
10\% better than that from the fit of events with $|d| > 0.3$.
This improvement is directly related to a better overall tagging power 
[Eq. (\ref{tagpower})] for the sum of subsamples as compared to the result
[Eq. (\ref{tagpower1})] for the sample with $|d|>0.3$.

\section{Systematic Uncertainties}
\label{sec:syserr} 

The systematic uncertainties are summarized in
Tables \ref{syste} and \ref{syste2}. 
Table \ref{syste} shows the
contributions to the systematic uncertainty in $\dmd$.
Table \ref{syste2} shows the corresponding contributions to the systematic
uncertainties in ${\cal D}(B^0)$.

\begin{table*}
\caption{Systematic uncertainties for $\dmd$.
\label{syste}
}
\begin{ruledtabular}
\begin{tabular}{|c|c|cc|cc|}
&Default&\doubcol{Variation}&\doubcol{$\dmd$ (ps$^{-1})$}\\\hline
&&(a)&(b)&(a)&(b)\\\hline
$\text{Br}(B^0 \to D^{*-}\mu^+\nu )$ & 5.44 & $-0.23$ & $0.23$ & ~0.002 & $-$0.002 \\
$\text{Br}(B \to D^*\pi\mu\nu X)$ & 1.07 & $-$0.17 & 0.17 & $-$0.0078 & ~0.0078 \\
$R^{**}$ & 0.35 & ~0 & 1.0 & ~0.0006 & $-$0.0012 \\
$B$ lifetimes & ~0.05022 & $-$0.00054 & ~0.00054 & ~0.0008 & $-$0.0008 \\
Resolution scale factor & \textemdash & 1.2 & 0.8 & ~0.0021 & $-$0.0021 \\
Alignment  & \textemdash & $-10 ~\mu$m & $+10 ~\mu$m & $-$0.004 & +0.004 \\
$K$-factor & \textemdash & $-$2\% & +2\% & ~0.0098 & $-$0.0094 \\
Efficiency & \textemdash & $-$12\% & +12\% & $-$0.0054 & ~0.0052 \\
Fraction $D^0$ in $D^*$ & 4\% & 3.15\% & 4.85\%   & $-$0.0020 & +0.0030 \\
\hline
Fit procedure & \multicolumn{5}{c|}{See below}\\ \hline
 Bin width & 2 MeV & 1.6 & 2.67 & ~0.0009 & ~0.0014 \\
Parameter $\mu_0$ & \textemdash & $-$$3\sigma$ & $3\sigma$ & $-$0.0001 & ~0.0001 \\
Parameter $\frac{\sigma_R+\sigma_L}{2}$ & \textemdash & $-$$3\sigma$ & $3\sigma$ & $-$0.0001 & \textemdash \\
Parameter $\frac{\sigma_R-\sigma_L}{\sigma_R+\sigma_L}$ & \textemdash & $-3\sigma$ & $3\sigma$ & $-$0.0001 & ~0.0001 \\
Parameter $\mu_1$ & \textemdash & $-$$3\sigma$ & $3\sigma$ & $-$0.0016 & ~0.0015 \\
Parameter $\frac{\sigma_1+\sigma_2}{2}$ & \textemdash & $-$$3\sigma$ & $3\sigma$ & $-$0.0006 & ~0.0006 \\
Parameter $R$ & \textemdash & $-$$3\sigma$ & $3\sigma$ & $-$.0005 & ~0.0004 \\
Parameter $(\mu_2-\mu_1)$ & \textemdash & $-$$3\sigma$ & $3\sigma$ & ~0.0006 & $-$0.0007 \\
Parameter $\frac{\sigma_1-\sigma_2}{\sigma_1+\sigma_2}$ & \textemdash & $-$$3\sigma$ & $3\sigma$ & \textemdash & \textemdash \\
\hline
Fit procedure&&\doubcol{Overall}&\doubcol{+0.0023}\\
&&\doubcol{}&\doubcol{$-$0.0019}\\
\hline
Total&&\doubcol{}&\doubcol{$\pm 0.0158$}\\
\end{tabular}
\end{ruledtabular}
\end{table*}

\begin{turnpage}
\begin{table*}
\scriptsize
\caption{Systematic uncertainties for ${\cal D}(B^0)$.\label{syste2}}
\begin{ruledtabular}
\begin{tabular}{|c|c|cc|cc|cc|cc|cc|cc|}
&&\doubcol{}&\doubcol{${\cal D}(B^0)$}&\doubcol{${\cal D}(B^0)$}&\doubcol{${\cal D}(B^0)$}&\doubcol{${\cal D}(B^0)$}&\doubcol{${\cal D}(B^0)$}\\
&Default&\doubcol{Variation}&\doubcol{$0.1<|d| \leq 0.2$}&\doubcol{$0.2 < |d| \leq 0.3$}&\doubcol{$0.3 < |d| \leq 0.45$}&\doubcol{$0.45 < |d| \leq 0.6$}&\doubcol{$0.6 < |d| \leq 1.0$}\\\hline
&&(a)&(b)&(a)&(b)&(a)&(b)&(a)&(b)&(a)&(b)&(a)&(b)\\\hline
$Br(B^0 \to D^{*-}\mu^+\nu )$ & 5.44 & $-$0.23 & ~0.23
& \textemdash & \textemdash & \textemdash & $-$0.001 & ~0.001 & \textemdash & ~0.001 & $-$0.001 & ~0.001 & $-$0.001 \\

$Br(B \to D^*\pi\mu\nu X)$ & 1.07 & $-$0.17 & 0.17
& ~0.0004 & $-$0.0004 & $-$0.0011 & ~0.0011 & $-$0.0019 & ~0.0021 & $-$0.0020 & ~0.0021 & $-$0.0008 & ~0.0028 \\

$R^{**}$ & 0.35 & ~0.0 & 1.0
& $-$0.0009 & ~0.0016 & $-$0.0027 & ~0.0048 & $-$0.0042 & ~0.0079 & $-$0.0057 & ~0.0105 & $-$0.0066 & ~0.0124 \\

$B$ lifetimes & ~0.05022 & $-$0.00054 & ~0.00054
& \textemdash & $-$0.0001 & ~0.0001 & $-$0.0002 & ~0.0003 & $-$0.0001 & ~0.0003 & $-$0.0003 & ~0.0014 & $-$0.0003 \\

Resolution function & \textemdash & $\times$1.2 & $\times$0.8
& ~0.0005 & $-$0.0006 & ~0.0010 & $-$0.0012 & ~0.0020 & $-$0.0021 & ~0.0024 & $-$0.0028 & ~0.0028 & $-$0.0032 \\

Alignment &  \textemdash & $-$10 $\mu m$ & 10 $\mu m$ 
& $-$0.004  &  0.004 & $-$0.004  & 0.004  & $-$0.004 & 0.004 & $-$0.004   & 0.004  &  $-$0.004 & 0.004  \\
$K$-Factor & \textemdash & $-$2\% & +2\%
& \textemdash & \textemdash & $-$0.0001 & \textemdash & \textemdash & ~0.0001 & $-$0.0001 & \textemdash & \textemdash & \textemdash \\

Efficiency & \textemdash & $-$12\% & +12\%
& ~0.0006 & $-$0.0007 & $-$0.0008 & ~0.0006 & $-$0.0012 & ~0.0011 & $-$0.0013 & ~0.0010 & $-$0.0021 & ~0.0019 \\

Fraction $D^0$ in $D^*$
& 4\% & 3.15\% & 4.85\% 
& \textemdash & ~0.0010 & $-$0.0010 & \textemdash & $-$0.0010 & ~0.0010 & $-$0.0010 & ~0.0010 & $-$0.0010 & ~0.0010\\

\hline Fit procedure & \multicolumn{13}{c|}{See split below}\\ \hline
 Bin width & 2 MeV & 1.6 & 2.67
& $-$0.0026 & ~0.0002 & $-$0.0024 & ~0.0014 & $-$0.0001 & ~0.0027 & ~0.0037 & ~0.0038 & ~0.0089 & ~0.0087 \\

Parameter $\mu_0$ & \textemdash & $-$$3\sigma$ & $3\sigma$
& $-$0.0003 & ~0.0002 & ~0.0001 & $-$0.0001 & ~0.0001 & ~0.0001 & $-$0.0002 & ~0.0001 & $-$0.0007 & ~0.0007 \\

Parameter $\frac{\sigma_R+\sigma_L}{2}$ & \textemdash & $-$$3\sigma$ & $3\sigma$
& ~0.0002 & $-$0.0002 & ~0.0001 & $-$0.0001 & ~0.0004 & $-$0.0003 & \textemdash & $-$0.0001 & $-$0.0002 & ~0.0001 \\

Parameter $\frac{\sigma_R-\sigma_L}{\sigma_R+\sigma_L}$ & \textemdash & $-$$3\sigma$ & $3\sigma$
& $-$0.0005 & ~0.0005 & ~0.0002 & $-$0.0001 & ~0.0002 & ~0.0001 & $-$0.0002 & ~0.0001 & $-$0.0015 & ~0.0011 \\

Parameter $\mu_1$ & \textemdash & -$3\sigma$ & $3\sigma$
& $-$0.0009 & ~0.0010 & $-$0.0017 & ~0.0018 & ~0.0023 & $-$0.0015 & ~0.0006 & $-$0.0005 & $-$0.0004 & $-$0.0004 \\

Parameter $\frac{\sigma_1+\sigma_2}{2}$ & \textemdash & $-$$3\sigma$ & $3\sigma$
& ~0.0008 & $-$0.0005 & ~0.0014 & $-$0.0009 & ~0.0037 & $-$0.0034 & $-$0.0013 & ~0.0017 & $-$0.0099 & ~0.0068 \\

Parameter $R$ & \textemdash & $-$$3\sigma$ & $3\sigma$
& ~0.0015 & $-$0.0011 & ~0.0029 & $-$0.0024 & ~0.0030 & $-$0.0027 & ~0.0013 & $-$0.0011 & $-$0.0046 & ~0.0035 \\

Parameter $(\mu_2-\mu_1)$ & \textemdash & $-$$3\sigma$ & $3\sigma$
& \textemdash & $-$0.0003 & ~0.0008 & $-$0.0011 & $-$0.0001 & ~0.0006 & $-$0.0003 & ~0.0002 & ~0.0008 & $-$0.0003 \\

Parameter $\frac{\sigma_1-\sigma_2}{\sigma_1+\sigma_2}$ & \textemdash & $-$$3\sigma$ & $3\sigma$
& $-$0.0001 & \textemdash & $-$0.0004 & ~0.0003 & ~0.0002 & $-$0.0002 & $-$0.0004 & ~0.0004 & $-$0.0006 & ~0.0010 \\

\hline
Fit procedure&&\doubcol{Overall}&\doubcol{+0.0021}&\doubcol{+0.0040}&\doubcol{+0.0060}&\doubcol{+0.0044}&\doubcol{+0.0119}\\
&&\doubcol{}&\doubcol{$-$0.0031}&\doubcol{$-$0.0041}&\doubcol{$-$0.0046}&\doubcol{$-$0.0019}&\doubcol{$-$0.0111}\\\hline
\hline
Total&&\doubcol{}&\doubcol{+.0049}&\doubcol{+.0077}&\doubcol{+.0111}&\doubcol{+.0125}&\doubcol{+.0182}\\
&&\doubcol{}&\doubcol{$-$0.0052}&\doubcol{$-$0.0066}&\doubcol{$-$0.0081}&\doubcol{$-$0.0081}&\doubcol{$-$0.0140}\\
\end{tabular}
\end{ruledtabular}
\end{table*}
\end{turnpage}

These uncertainties were obtained as follows:

\begin{itemize}
\item The $B$ meson branching fractions and lifetimes used in the fit of the
asymmetry were taken from Ref. \cite{pdg} and were varied by one standard deviation.

\item The VPDL resolution obtained in simulation was multiplied by
factors of 0.8 and 1.2.  These factors exceed the uncertainty 
in the difference of the resolution between data and simulation.

\item The variation of $K$-factors with the change in the $B$ momentum 
was neglected in this analysis. To check the impact of this assumption on the
final result, the computation of $K$-factors, was repeated without the cut 
on $p_T(\dzero)$ or by applying an additional cut on the $p_T$ of muon, 
$p_T > 4$ GeV/$c$. The change in the average values of the $K$-factors did not 
exceed 2\%, which was used as the estimate of the systematic uncertainty in their values. This uncertainty was propagated into the variation of $\dmd$
and tagging purity by repeating the fit with the $K$-factor distributions
shifted by 2\%.

\item The ratio of the reconstruction efficiencies in different $B$ meson decay channels
depends only on the kinematic properties of corresponding decays and
can therefore be reliably estimated in the simulation. The ISGW2
model \cite{isgw2} of semileptonic $B$ decays was used. The
uncertainty in the reconstruction efficiency, set at 12\%, was
estimated by varying the kinematic cuts on the $p_T$ of the muon and $D^0$
in a wide range. Changing the model describing semileptonic $B$
decay from ISGW2 to a HQET-motivated model \cite{hqet} produces a smaller
variation. The fit to the asymmetry was repeated with the efficiencies
to reconstruct the $B \to \mu^+ \nu \dststminus$ and $B \to \mu^+ \nu
\adststzero$ channels modified by $\pm 12\%$, and the difference was taken
as the systematic uncertainty from this source. 

\item The additional fraction of \dzero events contributing to the \dst sample
was estimated at $(4.00 \pm 0.85)\%$ (see Sec. \ref{sec:istbd}).
This variation was used to estimate the systematic uncertainty from this source.
As a cross-check, the number of \dst~ events was determined from the fit
of the mass difference $M(\dzero \pi) - M(\dzero)$ and the fit
of the flavor asymmetry was repeated. The measured value of 
$\dmd = 0.507 \pm 0.020$ ps$^{-1}$ is consistent with Eq. (\ref{result}).
\item We also investigated the systematic uncertainty in determining the
number of \dst~ and \dzero~ candidates in each VPDL bin. 

\begin{itemize}
\item The values of the parameters which had been fixed from the fit to ``all'' events,
were varied by $\pm 3\sigma$. 
\item The default bin width for the fits in the VPDL bin is 0.020 GeV. We
lowered the bin width to 0.016 GeV and increased the bin width to
0.027 GeV, and included the resulting variations in the systematic uncertainty.
\end{itemize}
\end{itemize}

\section{Conclusions}
We have performed a study of a likelihood-based opposite-side tagging
algorithm in $B^0$ and $B^+$ samples obtained with $\sim 1$ ~fb$^{-1}$ of
RunII Data.
The dilutions ${\cal D}(B^+)$ and ${\cal D}(B^0)$ were found to be 
the same within their statistical uncertainties. This result justifies
the application of the $B^0_d$ dilution to the $B^0_s$ mixing analysis.

Splitting the sample into bins according to the 
tagging variable $|d|$ and measuring the tagging power as 
the sum of the individual tagging powers of all bins, we obtained a tagging power of
$$\varepsilon 
{\cal D}^2=[{2.48 \pm 0.21~{\rm (stat.)}}^{+0.08}_{-0.06}~{\rm(syst)}]~ \%.$$

From the simultaneous fit to events in all $|d|$ bins we measured the 
mixing parameter:
$$\dmd = 0.506 \pm 0.020 ~{\rm (stat)} \pm 0.016~{\rm (syst)}~{\rm ps}^{-1},$$
which is in good agreement with the world average value of 
$\dmd = 0.502 \pm 0.007  {~\rm ps}^{-1} $ \cite{pdg}.
%


\begin{thebibliography}{99}
%
\bibitem[*]{kurca}
On leave from IEP SAS Kosice, Slovakia.
\bibitem[\dag]{voutilainen}
Visitor from Helsinki Institute of Physics, Helsinki, Finland.
%
\vskip 0.25cm
%
%
\bibitem{bdmix} H. Albrecht {\it et al.}, ARGUS Collaboration, Phys. Lett. B {\bf 192},
245 (1987);  M. Artuso {\it et al.}, CLEO Collaboration, Phys Rev. Lett. {\bf 62}, 2233 (1989).
\bibitem{sst} R. Akers {\it et al.}, OPAL Collaboration, Z. Phys. {\bf C 66}, 19 (1995).
\bibitem{pdg}
S. Eidelman {\it et al.}, Particle Data Group, Phys.\ Lett. B 
{\bf 592}, 1 (2004).
\bibitem{berger} E. Berger {\it et al.}, Phys. Rev. Lett. {\bf 86}, 4231 (2001).
\bibitem{dzero} V.M. Abazov {\it et al.}, D\O\ collaboration, 
Nucl. Instrum. and Methods {\bf A 565}, 463-537 (2006).
\bibitem{durham} S.~Catani {\it et al.}, Phys. Lett. B {\bf269}, 432 (1991).
\bibitem{pvreco} J. Abdallah {\it et al.}, DELPHI Collaboration,  
Eur. Phys. J. {\bf C32}, 185 (2004).
\bibitem{delphi} J. Abdallah {\it et al.}, DELPHI Collaboration, Euro. Phys. J. {\bf C35}, 35 (2004).
\bibitem{isgw2}
D. Scora and N. Isgur, Phys. Rev. D {\bf 52}, 2783 (1995).
\bibitem{hqet} M. Neubert, Phys. Rep. {\bf 245}, 259 (1994).
\bibitem{ALEPH}
D. Buskulic {\it et al.}, ALEPH Collaboration, Z. Phys. {\bf C 73}, 601 (1997).
\bibitem{DELPHI}
P. Abreu {\it et al.}, DELPHI Collaboration, Phys. Lett. B 
{\bf 475}, 407 (2000).
\bibitem{bsd0} V.~M.~Abazov {\it et al.}, D\O\ Collaboration, Phys. Rev. Lett. {\bf 97}, 021802 (2006).
\end{thebibliography}
%
%
We thank the staffs at Fermilab and collaborating institutions, 
and acknowledge support from the 
DOE and NSF (USA);
CEA and CNRS/IN2P3 (France);
FASI, Rosatom and RFBR (Russia);
CAPES, CNPq, FAPERJ, FAPESP and FUNDUNESP (Brazil);
DAE and DST (India);
Colciencias (Colombia);
CONACyT (Mexico);
KRF and KOSEF (Korea);
CONICET and UBACyT (Argentina);
FOM (The Netherlands);
PPARC (United Kingdom);
MSMT (Czech Republic);
CRC Program, CFI, NSERC and WestGrid Project (Canada);
BMBF and DFG (Germany);
SFI (Ireland);
The Swedish Research Council (Sweden);
Research Corporation;
Alexander von Humboldt Foundation;
and the Marie Curie Program.

\end{document}